\documentclass[structabstract]{aa}  

\usepackage{graphicx}
\usepackage{amssymb}
\usepackage{txfonts}
\usepackage{natbib}
\usepackage{color}
\bibpunct{(}{)}{;}{a}{}{,}

\begin{document}

\title{Circumstellar disks in binary star systems}
\subtitle{Models for $\gamma$~Cephei and $\alpha$~Centauri}

\author{Tobias~W.~A.~M\"uller
\and Wilhelm~Kley
}
\authorrunning{T. W. A. M\"uller and W. Kley}

\institute{
     Institut f\"ur Astronomie \& Astrophysik, 
     Universit\"at T\"ubingen,
     Auf der Morgenstelle 10, 72076 T\"ubingen, Germany \\
     \email{Tobias\_Mueller@twam.info}
}

\date{Received 4 October 2011 / Accepted 2 December 2011}

\abstract
{As of today, over 50 planetary systems have been discovered in binary star systems, some of which have binary separations that are smaller than $20$\,AU.
In these systems the gravitational forces from the binary have a strong influence on the evolution of the protoplanetary disk and hence the planet formation process.}
{We study the evolution of viscous and radiative circumstellar disks under the influence of a companion star. We focus on the eccentric $\gamma$~Cephei and
$\alpha$~Centauri system as examples and compare disk quantities such as disk eccentricity and precession rate to previous isothermal simulations.}
{We performed two-dimensional hydrodynamical simulations of the binary star systems under the assumption of coplanarity of the disk, host star and binary companion.
We used the grid-based, staggered mesh code \texttt{FARGO} with an additional energy equation to which we added radiative cooling based on opacity tables.}
{The eccentric binary companion perturbs the disk around the primary star periodically. Upon passing periastron, spirals arms are induced that wind from
the outer disk towards the star. In isothermal simulations this results in disk eccentricities up to $ e_\mathrm{disk} \approx 0.2 $, but in more realistic
radiative models we obtain much smaller eccentricities of about $ e_\mathrm{disk} \approx 0.04 - 0.06 $ with no real precession. Models with varying viscosity
and disk mass indicate that disks with less mass have lower temperatures and higher disk eccentricity.}
{The fairly high disk eccentricities, as indicated in previous isothermal disk simulations, implied a more difficult planet formation in the $\gamma$~Cephei
system caused by the enhanced collision velocities of planetesimals. We have shown that under more realistic conditions with radiative cooling the disk becomes
less eccentric and thus planet formation may be made easier. However, we estimate that the viscosity in the disk has to very low, with $\alpha \lesssim 0.001$,
because otherwise the disk's lifetime will be too short to allow planet formation to occur along the core instability scenario. We estimate that the periodic
heating of the disk in eccentric binaries will be observable in the mid-IR regime.}

\keywords{
	accretion, accretion disks --
	protoplanetary disks --
	hydrodynamics --
	radiative transfer --
	methods: numerical --
	planets and satellites: formation
	}

\maketitle

\section{Introduction}
Presently, about 50 planets are known to reside in binary stars systems. In all systems with solar-type
stars the planet orbits one of the stars while the secondary star acts as a perturber, i.e. these are in a so-called
S-type configuration \citep{1986A&A...167..379D}. Recently, planets seem to have been detected in
evolved binary star systems that are in a circumbinary (P-type) configuration  \citep[e.g.][]{2011A&A...526A..53B}.
The main observational characteristic
of the known planets in binary stars system have been summarized by \citet{2004A&A...417..353E,2006ApJ...646..523R,2007A&A...462..345D}.
Most planets are in binaries with very large separations with semi-major axis beyond 100\,AU,
in particular when detected by direct imaging. Observationally, there is evidence for fewer planets in binaries
with a separation of less than about 100\,AU \citep{2007A&A...474..273E}, in accordance with the expectation that binarity constitutes a challenge to
the planet formation process. Despite this, there are several systems
with a quite close binary separation such as $\gamma$~Cephei \citep{1988ApJ...331..902C},
Gliese\,86 \citep{2000A&A...354...99Q,2001A&A...370L...1E}, HD\,41004 \citep{2004A&A...426..695Z},
and HD\,196885 \citep{2008A&A...479..271C}.

As known from several theoretical studies, the presence of the secondary renders the planet formation process
more difficult than around single stars. Owing to the gravitational disturbance, the protoplanetary
disk is hotter and dynamically more excited such that the coagulation and growth process of planetesimals as well as
the gravitational instability process in the
early phase of planet formation are hindered \citep{2000ApJ...537L..65N,2004A&A...427.1097T,2006Icar..183..193T}.
This is particularly true for the mentioned closer binaries with an orbital separation of about $20$\,AU.
Hence this theoretical challenge has put these tighter binaries into the focus of studies on planet formation in binary stars
\citep{2007ApJ...660..807Q,2008MNRAS.386..973P,2008A&A...486..617K} and on their dynamical stability \citep{2004RMxAC..21..222D,2006ApJ...644..543H}.
The most famous example in this category is the $\gamma$~Cephei system, which has been known for over 20 years
to contain a protoplanet \citep{1988ApJ...331..902C,2003ApJ...599.1383H}. In recent years the orbital parameter of this
system have been updated  \citep{2007A&A...462..777N,2007ApJ...654.1095T} and the basic binary parameters are quoted in Table~\ref{tab:standardmodel}
(below) because $\gamma$~Cephei is our standard model. At the end of the paper we will briefly discuss the $\alpha$~Centauri system
as well.

Because planet formation occurs in disks, their dynamical structure is of crucial importance in estimating the 
efficiency of planetary growth processes. The prime effect of the secondary star is the truncation of the disk owing to
tidal torques \citep{1994ApJ...421..651A}. The truncation radius depends on the mass ratio of the binary, its
eccentricity and the viscosity in the disk. An additional effect, namely the excitation of eccentric modes in the disk,
which has been noticed previously in close binary stars \citep{1988MNRAS.232...35W,1991ApJ...381..259L,2008A&A...487..671K},
has recently drawn attention in studies of planet harboring binary stars as well \citep{2008MNRAS.386..973P,2008A&A...486..617K}.
In these studies it was shown that despite the high eccentricity of a $\gamma$~Cephei type binary the disk became eccentric with an average eccentricity
of about $e_{\rm disk} \approx 0.12$ and a coherent disk precession  \citep{2008A&A...486..617K}. However, this leaves the question of numerical effects \citep{2008MNRAS.386..973P}.
Subsequently the influence of self-gravity of the disk has been analyzed by \citet{2009A&A...508.1493M}, who concluded
that the inclusion of self-gravity leads to disks with lower eccentricity on average. 

Interestingly, there may be observational evidence for tidal interactions between a companion on an eccentric orbit and a
circumstellar disk around the primary star \citep{2010A&A...517A..16V}, as inferred from variable brightness on longer timescales.
The possibility to detect an eccentric disk through spectral line observations has been analyzed recently by \citet{2011A&A...528A..93R}.

In all simulations mentioned above, the disk has typically been modeled using a fixed radial temperature distribution.
This simplifies the numerics such that no energy equation has to be solved but of course it lacks physical reality. In this work
we will extend previous studies and consider a much improved treatment of the energy equation. We will do so within
the two-dimensional, flat-disk approximation following \citet{2003ApJ...599..548D} and \citet{2008A&A...487L...9K}.
Here, the viscous heating of the disk and the radiative losses are taken into account. We will study the structure and dynamical evolution
of disks with different masses and binary parameter, and will analyze the influence of numerical aspects, in particular boundary conditions.
Our first target of interest will be the well-studied system $\gamma$~Cephei and we will present results on the $\alpha$~Centauri system subsequently.

\section{Model setup}
We assumed that the complete system of primary star, circumstellar disk and secondary star is coplanar.
Consequently, in our calculations we assumed a flat two-dimensional disk orbiting the primary. 
The disk is modeled hydrodynamically, under the assumption that the action of the turbulence can be described by a
standard viscous stress tensor.

\begin{table*}[hbt]
	\centering
	\renewcommand\arraystretch{1.2}
	\begin{tabular}{lcccccc}
	\hline
        \# & regime & $ \kappa_0 $ [$\mathrm{cm}^2/\mathrm{g}$] & $a$ & $b$ & $T_\mathrm{min}$ [K] & $ T_\mathrm{max} $ [K] \\
	\hline\hline
	1 & Ice grains & $ 2 \times 10^{-4} $ & $ 0 $ & $ 2 $ & $ 0 $ & $ 170 $ \\
	2 & Sublimation of ice grains & $ 2 \times 10^{16} $ & $ 0 $ & $ -7 $ & $ 170 $ & $ 210 $ \\
	3 & Dust grains & $ 5 \times 10^{-3} $ & $ 0 $ & $ 1 $ & $ 210 $ & $ 4.6 \times 10^{3} \rho^{\frac{1}{15}} $ \\
	4 & Sublimation of dust grains & $ 2 \times 10^{34} $ & $ \frac{2}{3} $ & $ -9 $ & $ 4.6 \times 10^{3} \rho^{\frac{1}{15}} $ & $ 3000 $ \\
	5 & Molecules & $ 2 \times 10^{-8} $ & $ \frac{2}{3} $ & $ 3 $ & $ 3000 $ & $ 1.1 \times 10^{4} \rho^{\frac{1}{21}} $ \\
	6 & Hydrogen scattering& $ 1 \times 10^{-36} $ & $ \frac{1}{3} $ & $ 10 $ & $ 1.1 \times 10^{4} \rho^{\frac{1}{21}} $ & $ 3 \times 10^{4} \rho^{\frac{4}{75}} $ \\
	7 & bound-free \& free-free & $ 1.5 \times 10^{20} $ & $ 1 $ & $ -\frac	{5}{2} $ & $ 3 \times 10^{4} \rho^{\frac{4}{75}} $ & ---  \\ \hline
	\end{tabular}
	\caption{
		\label{tab:opacity}
		Details of the various opacity regimes by type, showing the transition temperature and the constants $ \kappa_0$, $ a$ and $b$. All values are quoted in cgs units. See \citet{1985prpl.conf..981L} for more details.
	}
\end{table*}

\subsection{Physics and equations}
We used cylindrical coordinates $ (r, \varphi, z)$ centered on the primary star where the disk lies in the equatorial,
$z = 0$ plane. 
Because our model is two-dimensional $(r, \varphi)$, we solved the vertically integrated versions of the hydrodynamical equations. In this approximation the continuity equation is 
\begin{eqnarray}
	\label{mass_eq}
	\frac{\partial \Sigma}{\partial t} + \nabla \cdot (\Sigma \vec{v}) &=& 0\,,
\end{eqnarray}
where $ \vec{v} = (v_r, v_\varphi) = (v, \Omega r) $ is the velocity,  and $ \Sigma = \int_{-\infty}^{\infty} \rho \,dz $ the surface density. As indicated, in the following we will also use $v$ and $r \Omega$ for the radial and orbital velocity, respectively.
The vertically integrated equation of radial motion is then
\begin{eqnarray}
	\label{momentum_eq}
	\frac{\partial (\Sigma v)}{\partial t} + \nabla \cdot (\Sigma v \vec{v}) &=& \Sigma r \Omega^2 - \frac{\partial p}{\partial r} - \Sigma \frac{\partial \Psi}{\partial r} + f_r \,,
\end{eqnarray}
and for the azimuthal component
\begin{eqnarray}
	\label{angular_momentum_eq}
	\frac{\partial (\Sigma r^2 \Omega)}{\partial t} + \nabla \cdot (\Sigma r^2 \Omega \vec{v}) &=& - \frac{\partial p}{\partial \varphi} - \Sigma \frac{\partial \Psi}{\partial \varphi} + r^2 f_\varphi\,. 
\end{eqnarray}
Here $p$ is the vertically integrated pressure, $ \Psi $ the gravitational potential of both stars, and $f_r $ and $ f_\varphi $ describe the radial and azimuthal forces due to the disk viscosity \citep[][]{2002A&A...387..605M}.
Owing to the motion of the primary star around the center of mass of the binary, the coordinate system is non-inertial,
and indirect terms were included in the equations of motion to account for this. These are included in the potential $ \Psi $. 
The gravitational influence of the disk on the binary is neglected, and the disk in non-self-gravitating. 
The vertically integrated energy equation reads
\begin{eqnarray}
	\label{energy_eq}
	\frac{\partial e}{\partial t} + \nabla (e \vec{v})  &=& - p \nabla \cdot \vec{v} + Q_+ - Q_-\,,
\end{eqnarray}
where $ e $ is the internal energy density, $ Q_+ $ the heating source term and $ Q_- $ the cooling source term. To obtain a fully determined system, we additionally used the ideal gas law
\begin{eqnarray}
	p = \mathcal{R} \Sigma T = ( \gamma - 1 ) e\,, 
\end{eqnarray}
where $ T $ is the temperature in the midplane of the disk, $ \gamma $ the adiabatic index and $\mathcal{R} $ the universal gas constant divided by the mean molecular mass,
which can be calculated by $ \mathcal{R} = k_\mathrm{B} / (\mu m_u) $, where $ k_\mathrm{B} $ is the Boltzmann constant, $ \mu $ the mean molecular weight and $ m_u $ the unified atomic mass unit.

The adiabatic sound speed $ c_\mathrm{s} $ within the disk is then given as
\begin{eqnarray}
	c_\mathrm{s} = \sqrt{\gamma \frac{p}{\Sigma}} = \sqrt{\gamma}\,c_\mathrm{s,iso}\,,
\end{eqnarray}
where $ c_\mathrm{s,iso} =  \sqrt{p/\Sigma} $ is the isothermal sound speed. The vertical pressure scale height $ H $ is then 
\begin{eqnarray}
	H = \frac{c_\mathrm{s,iso}}{\Omega_K} = \frac{c_\mathrm{s}}{\sqrt{\gamma}\,\Omega_\mathrm{K}} = \frac{c_\mathrm{s}}{\sqrt{\gamma}\,v_\mathrm{K}} r = h r\,,
\end{eqnarray}
where $ \Omega_\mathrm{K} $ denotes the Keplerian angular velocity around the primary and $h$ the aspect ratio.

For the heating term $ Q_+ $ we assumed that this is solely given by viscous dissipation, and it then is given by
\begin{eqnarray}
	Q_+ & = & \frac{1}{2 \nu \Sigma} \left[ \sigma_{rr}^2 + 2 \sigma_{r \varphi}^2 + \sigma_{\varphi \varphi}^2 \right] + \frac{2 \nu \Sigma}{9} (\nabla \vec{v})^2\,,
\end{eqnarray}
where $ \nu $ is the kinematic viscosity and $ \sigma $ denotes the viscous stress tensor, to be written in polar coordinates. The viscosity $ \nu $ is given by $ \nu = \alpha c_\mathrm{s} H $ \citep{1973A&A....24..337S}.

The cooling term $ Q_- $ describes the radiative losses from the lower and upper disk surface, which can be written as
\begin{eqnarray}
	Q_- = 2 \sigma_\mathrm{R} \frac{T^4}{\tau_\mathrm{eff}} \,,
\end{eqnarray}
where $ \sigma_\mathrm{R} $ is the Stefan-Boltzmann constant and $ \tau_\mathrm{eff} $ an effective optical depth.
We followed the approach of \citet{2008A&A...487L...9K} and write, according to \citet{1990ApJ...351..632H},
\begin{eqnarray}
	\tau_\mathrm{eff} = \frac{3}{8} \tau + \frac{\sqrt{3}}{4} + \frac{1}{4 \tau + \tau_\mathrm{min}}\,.
\end{eqnarray}
The optical depth follows from $\tau = \int \rho \kappa dz$, that can be approximated by
$\tau \approx \rho \kappa H$ where $\rho$ and $\kappa(\rho, T)$ are evaluated at the disk's midplane.
The vertical density profile of a disk is approximately given by a Gaussian and hence 
\begin{equation}
	\label{eq:sigma}
	\Sigma = \sqrt{2 \pi} \, \rho H\,.
\end{equation}
To account for the drop of opacity with vertical height we introduced a correction factor $ c_1 $ and write finally
\begin{eqnarray}
	\tau = \frac{c_1}{\sqrt{2 \pi}} \, \kappa \Sigma\,.
\end{eqnarray}
The constant $ c_1 = \frac{1}{2} $ is obtained by comparing two-dimensional disk models with the fully three-dimensional calculations as presented
in \citet{2009A&A...506..971K}. For the Rosseland mean opacity $ \kappa $ we adopted power-law dependencies on temperature and density described by 
\citet{1985prpl.conf..981L} and \citet{1994ApJ...427..987B}, where
\begin{eqnarray}
	\kappa = \kappa_0 \rho^a T^b
\end{eqnarray}
for various opacity regimes. Each opacity regime is described by a minimum temperature $ T_\mathrm{min} $ and maximum temperature $ T_\mathrm{max} $, which depends on the density $ \rho $.
Table~\ref{tab:opacity} lists the constants $ \kappa_0 $, $ a$ and $ b $ for each regime of
the \citeauthor{1985prpl.conf..981L} model. The temperature and density are taken from the midplane, where the density $\rho$ is obtained from Eq.~(\ref{eq:sigma}).

In our model we did not consider presently any radiation transport within the disk plane.
 This contribution is potentially important when strong gradients in temperature and density occur.
 In our situation this may be the case around the periastron phase of the binary.
 However, in our simulations the observed contrast did not seem strong enough and we did not expect a large impact on the evolution. 
 In subsequent studies we plan to investigate this question further.

Because we are interested in the global evolution of the disk, we measured the disk eccentricity $ e_\mathrm{disk} $ 
and the disk periastron $ \varpi_\mathrm{disk} $ by first calculating for each grid cell the
eccentricity vector $ \vec{e} $, which is defined by
\begin{eqnarray}
	\vec{e} &= \frac{\vec{v} \times \vec{j}}{G M} - \frac{\vec{r}}{|\vec{r}|} \,,
\end{eqnarray}
where $ \vec{j} = \vec{r} \times \vec{v} $ is the specific angular momentum, $ G $ the gravitational constant, $ M $ the total mass and $ \vec{r} $ the relative vector. In our two-dimensional case the specific angular
momentum only has a component in $ z $ direction and therefore the eccentricity $ e $ and the longitude of periastron $ \varpi $ follow as
\begin{eqnarray}
	e &=& | \vec{e} | = \sqrt{ e_x^2 + e_y^2 } \\
	\varpi &=& \mathrm{atan2}(e_y, e_x)\,.
\end{eqnarray}
The global disk eccentricity $ e_\mathrm{disk} $ and disk periastron $ \varpi_\mathrm{disk} $ is then calculated by a mass-weighted average
over the whole disk,
\begin{eqnarray}
	e_\mathrm{disk} &=&  \left[ \int_{r_\mathrm{min}}^{r_\mathrm{max}} \int_0^{2 \pi} \Sigma e\,r \,d\varphi\,dr \right] \times \left[ \int_{r_\mathrm{min}}^{r_\mathrm{max}} \int_0^{2 \pi} \Sigma\,r \,d\varphi\,dr \right]^{-1} \\
	\varpi_\mathrm{disk} &=& \left[ \int_{r_\mathrm{min}}^{r_\mathrm{max}}  \int_0^{2 \pi} \Sigma \varpi\,r\,d\varphi\,dr \right] \times \left[\int_{r_\mathrm{min}}^{r_\mathrm{max}} \int_0^{2 \pi}  \Sigma\,r\,d\varphi\,dr \right]^{-1}
\end{eqnarray}
where the integrals are evaluated by summation over all grid cells.

\begin{table}
	\centering
	\renewcommand\arraystretch{1.2}
	\begin{tabular}{ll}
		\hline 
		Primary star mass ($ M_\mathrm{primary} $) & $ 1.4\,M_{\sun} $ \\
		Secondary star mass ($ M_\mathrm{secondary} $) & $ 0.4\,M_{\sun} $ \\
		Binary semi-major axis ($ a $) & $ 20\,\mathrm{AU} $ \\
		Binary eccentricity ($ e_\mathrm{bin} $) & $ 0.4 $ \\ 
		Binary orbital period ($ P_\mathrm{bin} $) & $ 66.6637 \,\mathrm{a} $ \\
              \hline
		Disk mass ($ M_\mathrm{disk} $) & $ 0.01\,M_{\sun} $ \\
		Viscosity ($ \alpha $) & $ 0.01 $ \\
		Adiabatic index ($ \gamma $) & $ 7/5 $ \\
		Mean-molecular weight ($ \mu $) & $ 2.35 $ \\
              \hline
		Initial density profile ($ \Sigma $) & $ \propto r^{-1} $ \\
		Initial temperature profile ($ T $) & $ \propto r^{-1} $ \\
		Initial disk aspect ratio ($ H/r $) & $ 0.05 $ \\
              \hline
		Grid ($N_r \times N_\varphi$) & $ 256 \times 574 $ \\
		Computational domain ($ R_\mathrm{min} $ -- $ R_\mathrm{max} $) & $ 0.5$ -- $8\,\mathrm{AU} $ \\
		\hline
	\end{tabular}
	\caption{
		\label{tab:standardmodel}
		Parameters of the standard model. The top entries refer to the fixed binary parameter. The following give the
		disk properties, and then the initial disk setup and the computational parameter are given.
	}
\end{table}
\subsection{Numerical considerations}

The simulations were performed using the \texttt{FARGO} code \citep{2000A&AS..141..165M} with modifications from \citet{baruteauthesis}. The numerical method used in \texttt{FARGO} is a staggered mesh finite difference method. It uses operator splitting
and a first-order integrator to update to velocities with the source terms (potential and pressure gradients, viscous and centrifugal accelerations). The advective terms are treated by a second-order
upwind algorithm \citep{1977JCoPh..23..276V}. To speed up calculations the code uses the FARGO algorithm \citep{2000A&AS..141..165M}. The algorithmic details of the \texttt{FARGO} code have been described
in \citet{2000A&AS..141..165M}. Because the \texttt{FARGO} code is based on \texttt{ZEUS-2D}, the basic techniques are described in \citet{1992ApJS...80..753S}.

The position of the secondary star is calculated by a fifth-order Runge-Kutta algorithm. To smooth shocks we used the artificial viscosity described by \citet{1992ApJS...80..753S} in our simulations.

To test the code we checked that we obtained the same results when using a corotating coordinate system. In addition, several test calculations were made using the \texttt{RH2D} code \citep{1989A&A...208...98K,1999MNRAS.303..696K} 
to assure that we implemented everything correctly.

To avoid numerical problems, we implemented a surface density floor of $ \Sigma_\mathrm{floor} = 10^{-7} \times \Sigma_0 $ where $ \Sigma_0 = \Sigma(1\,\mathrm{AU})|_{t=0} $ and a temperature
floor of $ T_\mathrm{floor} = 3$\,K, which is about the temperature of the cosmic background radiation. 
In addition to the physical motivation for the floor, very low temperatures only occur in the very outer parts
of the disk (see e.g. Fig.~\ref{fig:gc-times-sig-temp-ecc}) where there is only very little mass that influences the dynamics of
the disk. For the same reason, these cells do not contribute to the mass weighted average of the disk eccentricity.
We performed test simulations with a very low temperature floor and obtained identical results.

\begin{figure*}[htb]
	\centering
	\includegraphics[width=\textwidth]{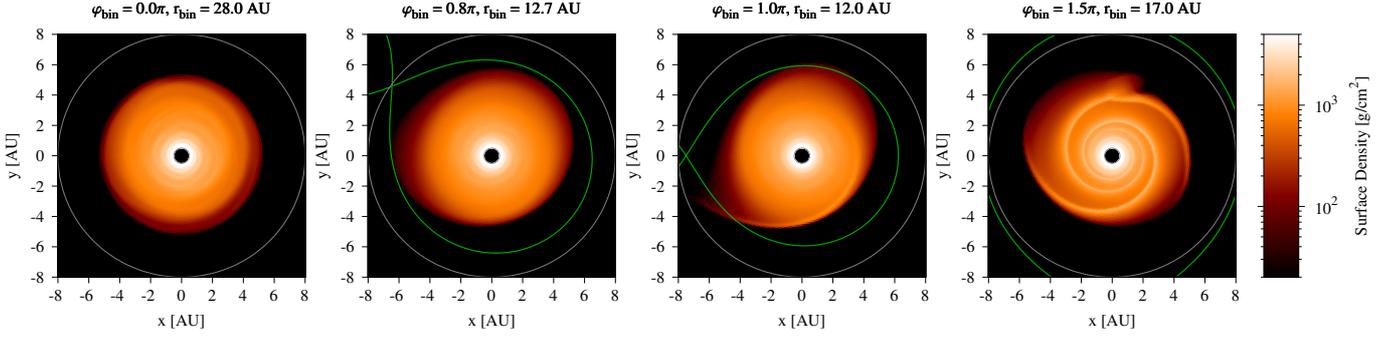} 
	\caption{
		\label{fig:gc-std-density}
		Evolution of the disk within one binary period for the {fully radiative standard model}. 
		The top labels in the four panels state the current position of the binary ($r_\mathrm{bin}$, $\varphi_\mathrm{bin}$) in polar coordinates.
		In the \textit{first} panel the binary companion is at the apastron and has less influence on the disk. The $ L_1 $ point in this
		configuration is at about $ 17.5$\,AU and therefore far outside the computational domain (gray circles).
		In the \textit{second} panel the binary has reduced its distance to about $ 12.8 $\,AU and the Roche lobe (green curve) has shrunken dramatically
                and is now entirely within the computational domain.
		The \textit{third} panel shows the disk with the binary at periastron. Some of the material in the disk is now outside the Roche lobe
                and might be lost from the system.
		In the \textit{last} panel the binary's separation increases again, which makes the Roche lobe grow such that it engulfs the disk entirely.
                The strong tidal forces near periastron induce spiral waves in the disk that will be damped out until the binary is at apastron.
	}
\end{figure*}

For the simulations we typically used two boundary conditions. 
The open boundary condition allows outflow, but no inflow of material. It is implemented as a zero-gradient outflow condition,
which reads at $R_\mathrm{min}$ as 
\begin{eqnarray*}
	\Sigma_{0j} &=& \Sigma_{1j} \\
	e_{0j} &=& e_{1j} \\
	v_{0j} &=& v_{1j} = 0 \textnormal{ if } v_{2j} > 0 \textnormal{ and } v_{2j} \textnormal{, otherwise,}
\end{eqnarray*}
where the index $0$ denotes the first inner radial ghost cells, e.g. $\Sigma_{1j}$ is the first active cell in radial and the 
$ j$-th cell in azimuthal direction. 
Alternatively, the reflecting boundary condition denies exchange of quantities with the system as is implemented at $R_\mathrm{min}$ as
\begin{eqnarray*}
	\Sigma_{0j} &=& \Sigma_{1j} \\
	e_{0j} &=& e_{1j} \\
	v_{0j} &=& v_{2j} \\
	v_{1j} &=& 0
\end{eqnarray*}
and similarly at $R_\mathrm{max}$.
For the outer radial boundary we always used the standard outflow condition, while at $R_\mathrm{min}$ we typically
used the reflecting condition. Below, we will investigate a possible influence of the numerical inner boundary condition.
 
\section{The standard model}
\label{sec:standarmodel}

For our simulations we constructed a standard reference model using the physical parameter of the $\gamma$~Cephei system. 
The first entries of Table \ref{tab:standardmodel} list the orbital parameter of this standard model based on the
observations of \citet{2007A&A...462..777N} and \citet{2007ApJ...654.1095T}.

The disk parameters where chosen to be in a typical range for a possible disk around $\gamma$~Cephei A, with a total mass
that would allow for the formation of the observed planet of about $1.85 M_\mathrm{Jup} $ \citep{2011AIPC.1331...88E} in the system.
Because the tidal forces of the secondary limits the radial extent of the disk to about $6$\,AU \citep{1994ApJ...421..651A}
we chose a slightly larger extent of the computational domain to $8$\,AU. This reduces artificial boundary effects.
The initial density distribution is limited to $6$\,AU.
For the outer boundary condition we always used the zero-gradient outflow boundary condition so that the material
that is accelerated during periastron may leave the system freely. The inner boundary in the standard model is typically reflecting. 
The resolution of the logarithmic grid is trimmed to have quadratic cells ($ r \Delta \varphi / \Delta r \approx 1 $).

After the initial setup at time zero, the models have to run for several binary orbits until a quasistationary state has been reached,
see also \citet{2008A&A...486..617K}. This equilibration process typically takes about 15 binary periods, i.e. 1000 years for
fully radiative models. 
Afterward the disk cycles through several states during one orbital period of the binary displayed in Fig.~\ref{fig:gc-std-density}.
When the binary is at apastron, the the disk is very axisymmetric, but when the binary moves toward its periastron, it starts to perturb the disk.
After the binary passes its periastron, two strong tidal spiral arms develop within the disk and wind themselves to the center of the disk.
Before the binary reaches its apastron, the disk spiral arms disappear. Periastron passages are also the time when most of the disk mass is lost.
During the first periapse the disk looses $4.8\,\%$ of its mass. This decreases during the next ten periapsis passes to about $0.1$\,\%. 
At the end of the simulation after $ 200 $ binary orbits, the disk mass has reduced to $0.0073\,M_{\sun}$, which is a loss of about $20\,\%$.

\subsection{Isothermal runs}
\label{sec:isothermal}
\begin{figure}
	\centering
	\includegraphics[width=\columnwidth]{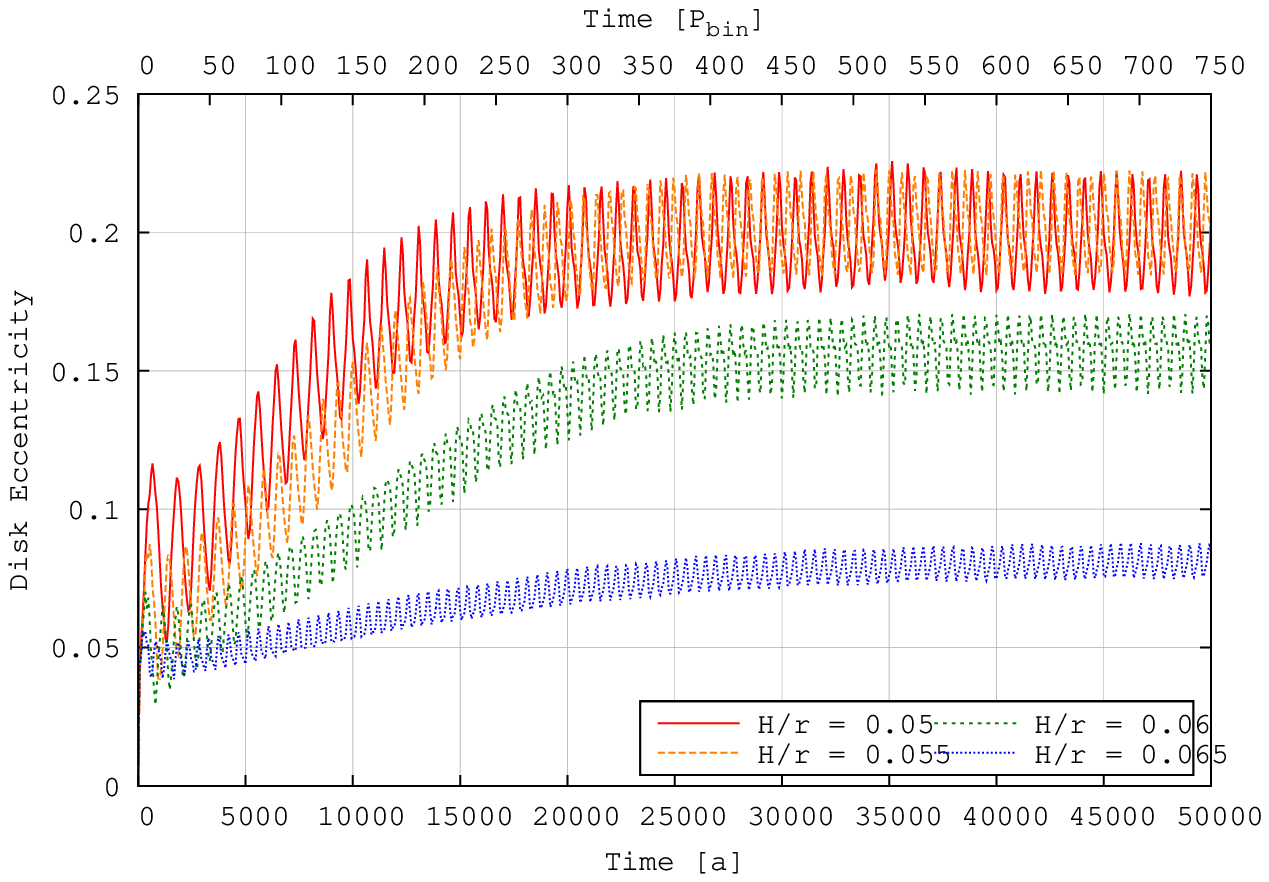} 
	\includegraphics[width=\columnwidth]{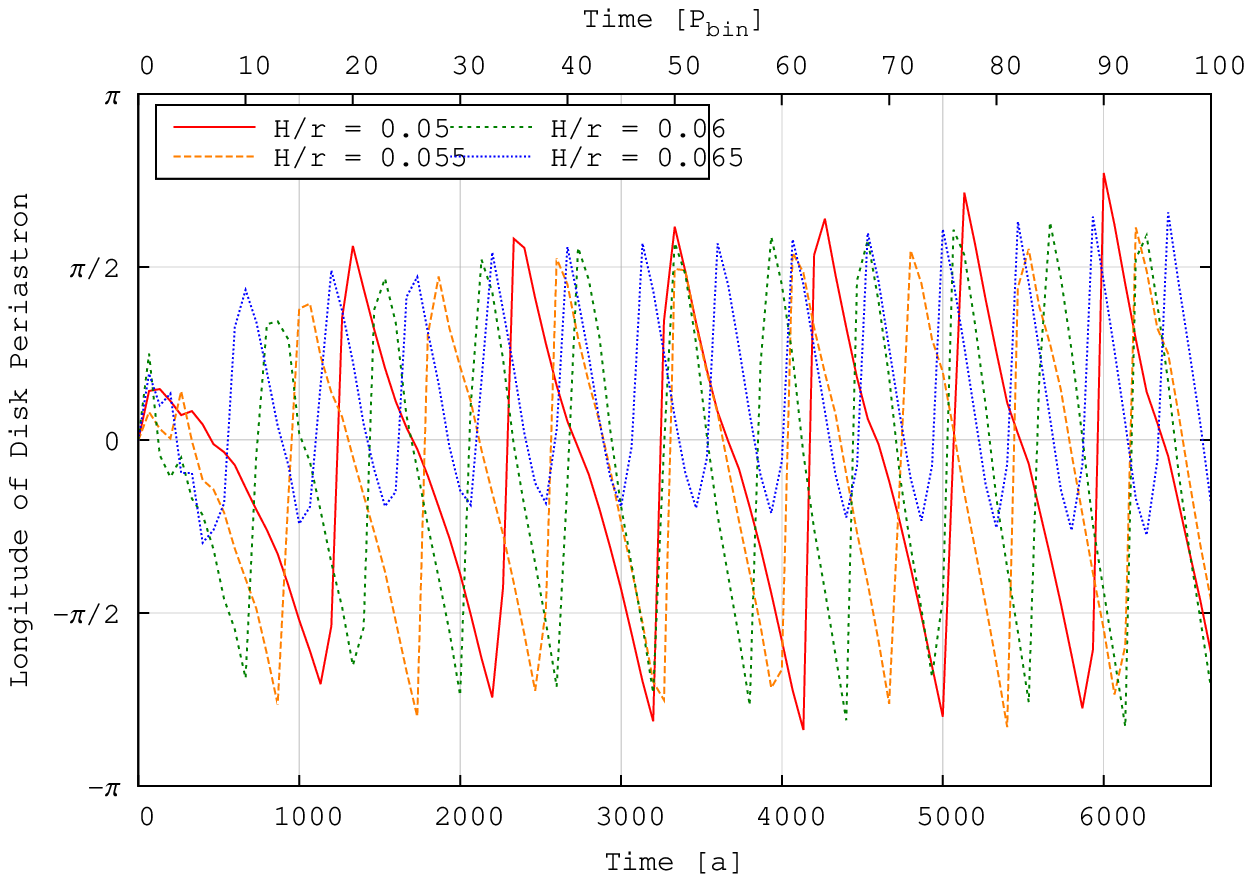}
	\caption{
		\label{fig:gc-iso}
		Global mass-weighted disk eccentricity (\textit{top}) and disk periastron (\textit{bottom}),
            both sampled at the binary's apastron, for \textit{isothermal} simulations of the standard model with different aspect ratios $H/r $.
		The disk eccentricity needs quite a long time to reach a quasi-equilibrium and settles at around $0.2$ for an $H/r$ of $0.05$ and $0.055$. High values of $H/r$ result in a 
		lower disk eccentricity. Note that in the lower panel a short time span is displayed for clarity. }
\end{figure}

To check our results and compare them with previous results, we first performed simulations where we kept the initial temperature stratification.
In Fig.~\ref{fig:gc-iso} we present results for constant $H/r$ disks where we did not solve the energy equation. Shown are results
for four different values of $H/r$, which were chosen according to our fully radiative results (see below). 
The disk reaches an equilibrium state after many (typically several 100) binary orbits, and the magnitude of the final
average eccentricity depends on the temperature of the disk. In particular, for
the disks with an  $H/r$ of about $0.05$ the disk eccentricity settles at fairly high values of about $0.2$. 
This seems to be higher than in previous simulations performed by \citet{2008A&A...486..617K}, but the parameters used 
for the $\gamma$~Cephei system, in particular the mass ratio $q = M_\mathrm{secondary} / M_\mathrm{primary} $
of the host star and the binary companion (their $q= 0.24$ instead of our $q = 0.29$), and the binary eccentricity $e_\mathrm{bin}$ ($ e_\mathrm{bin} = 0.36 $ instead of $ e_\mathrm{bin} = 0.4 $) are different.
Using the (older) system parameters as given by \citeauthor{2008A&A...487..671K}, we were able to reproduce the results nicely (e.g. their Fig.~4).

In these isothermal runs it is clear that the two thicker, hotter disks have significantly lower mean eccentricities.
The disk with $H/r=0.065$ does not even display a coherent disk precession any more (Fig.~\ref{fig:gc-iso}, bottom panel). 
This drop of $e_{\rm disk}$ with increasing $H/r$ is a consequence of the reduction of the eccentricity growth rate for thicker disks.
The growth rate of the disk eccentricity is the lowest for the hottest disk with $H/r = 0.065 $ and increases with
decreasing $H/r$ (Fig.~\ref{fig:gc-iso}, top panel), a finding that perfectly agrees with \citet{2008A&A...487..671K}.
The hotter disks need therefore much more time to reach an equilibrium state. 
As the eccentric binary damps the growth of disk eccentricity by tidal interaction (see below), the
hotter disks settle into an equilibrium state that has a lower $e_{\rm disk}$.

After about 400 binary periods, i.e. 25000 years, the time average of the disk eccentricity $ \langle e_\mathrm{disk} \rangle $ reaches a constant value and the disks precess with a
nearly constant precession rate $\dot{\varpi}_\mathrm{disk}$. The values of $ \langle e_\mathrm{disk} \rangle $  and $ \dot{\varpi}_\mathrm{disk} $ for the four different
values of $H/r$ are given in Table \ref{tab:isothermal}.

\begin{table}[hbt]
	\centering
	\renewcommand\arraystretch{1.2}
	\begin{tabular}{c|cc}
		$ H/r $ & $  \langle e_\mathrm{disk} \rangle $ & $ \dot{\varpi}_\mathrm{disk} $ [$P_\mathrm{bin}^{-1}$] \\ \hline
		$0.05$  & $ 0.19803 \pm 0.00085 $ & $ -0.0844 \pm 0.0018 $ \\
		$0.055$ & $ 0.20200 \pm 0.00076 $ & $ -0.1082 \pm 0.0017 $ \\
		$0.06$  & $ 0.15800 \pm 0.00054 $ & $ -0.1172 \pm 0.0063 $ \\
		$0.065$ & $ 0.08004 \pm 0.00031 $ & $ -0.1334 \pm 0.0025 $ \\
	\end{tabular}
	\caption{\label{tab:isothermal}
		Time average of the disk eccentricity $ \langle e_\mathrm{disk} \rangle $ and precession rate $ \dot{\varpi}_\mathrm{disk} $ for the four different \textit{isothermal} models.
	}
\end{table}

\begin{figure}
	\centering
	\includegraphics[width=\columnwidth]{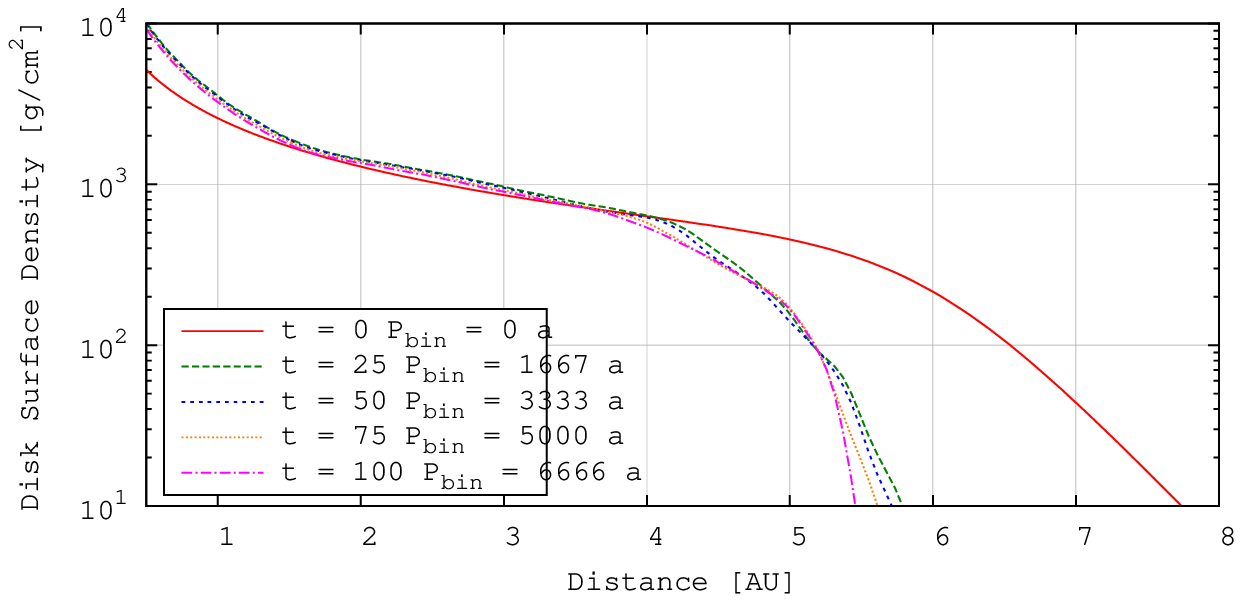} 
	\includegraphics[width=\columnwidth]{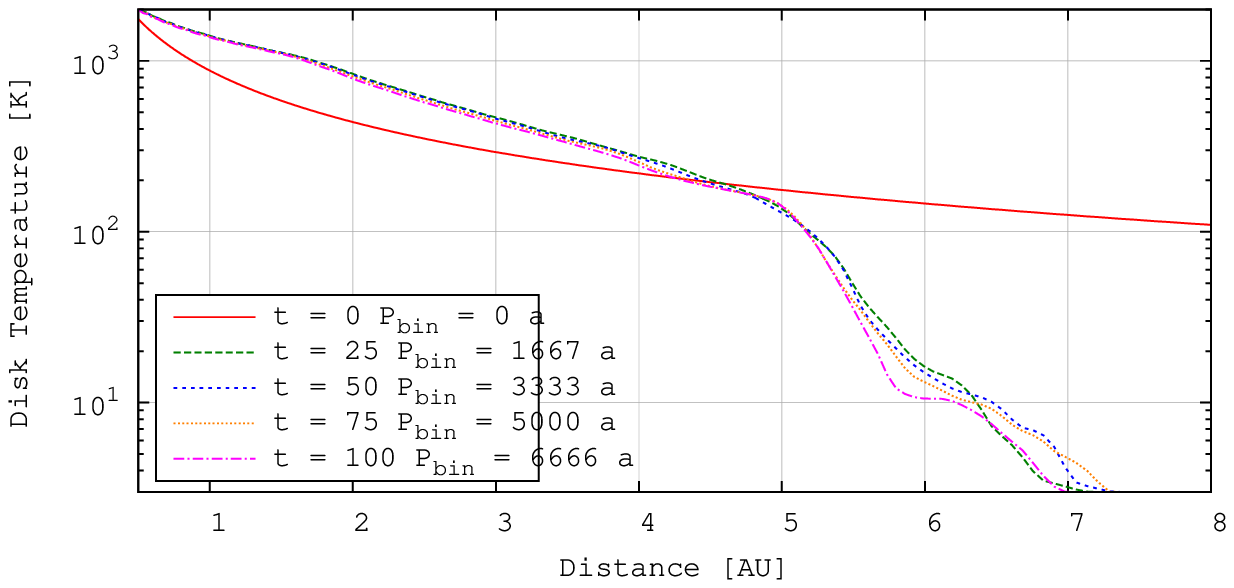} 
	\includegraphics[width=\columnwidth]{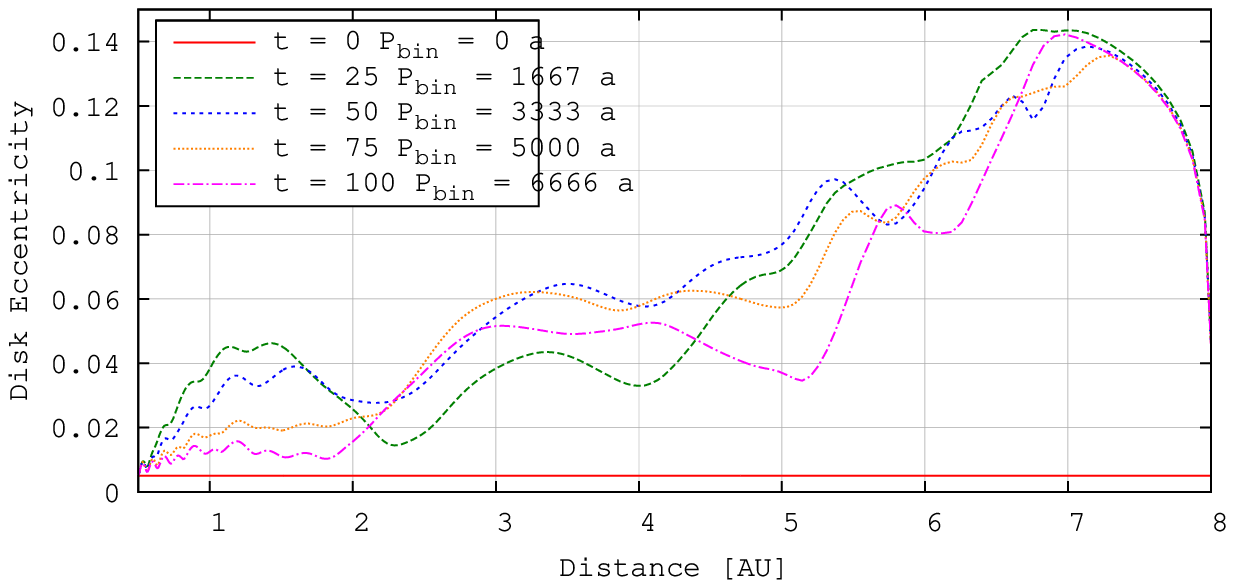} 
	\caption{
		\label{fig:gc-times-sig-temp-ecc}
		Radial dependency of surface density (\textit{top}), midplane temperature (\textit{middle}) and disk eccentricity (\textit{bottom}) of the \textit{fully radiative} standard model for different timestamps.
		The disk ranges from $ 0.5$\,AU to about $5.5$--$6$\,AU due to tidal forces of the companion. At $t = 0$\,years the disk is not fully Keplerian and thus $e_\mathrm{disk} > 0$.
	}
\end{figure}

\subsection{Fully radiative disks: Structure and dynamics}
\label{sec:structureanddynamics}

\begin{figure}
	\centering
	\includegraphics[width=\columnwidth]{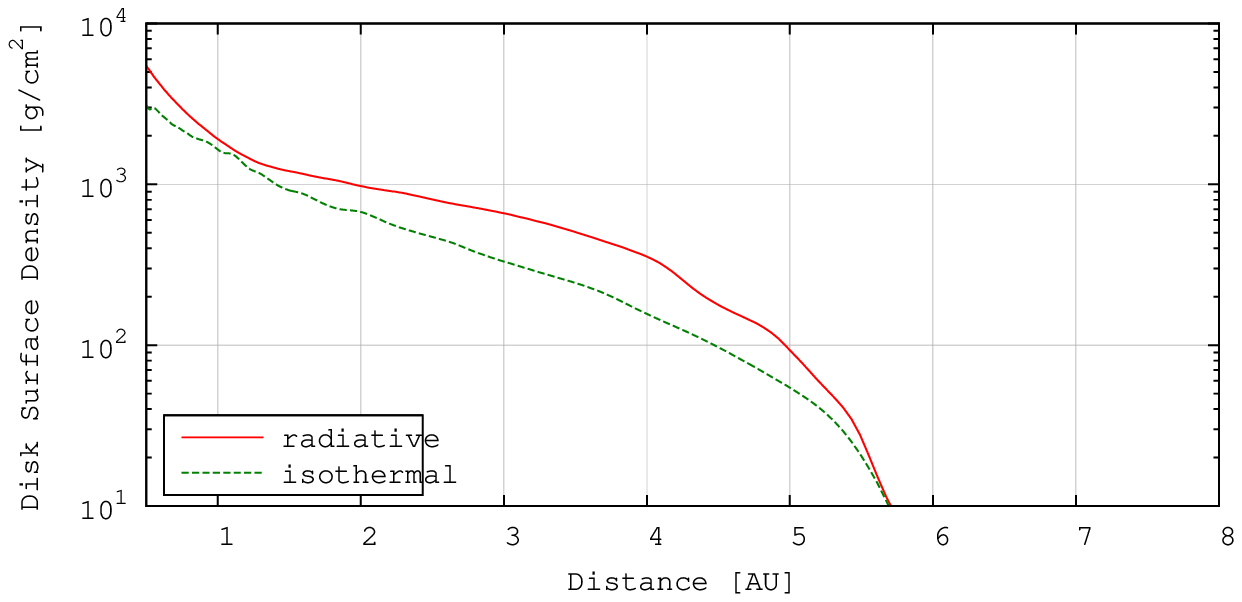} 
	\includegraphics[width=\columnwidth]{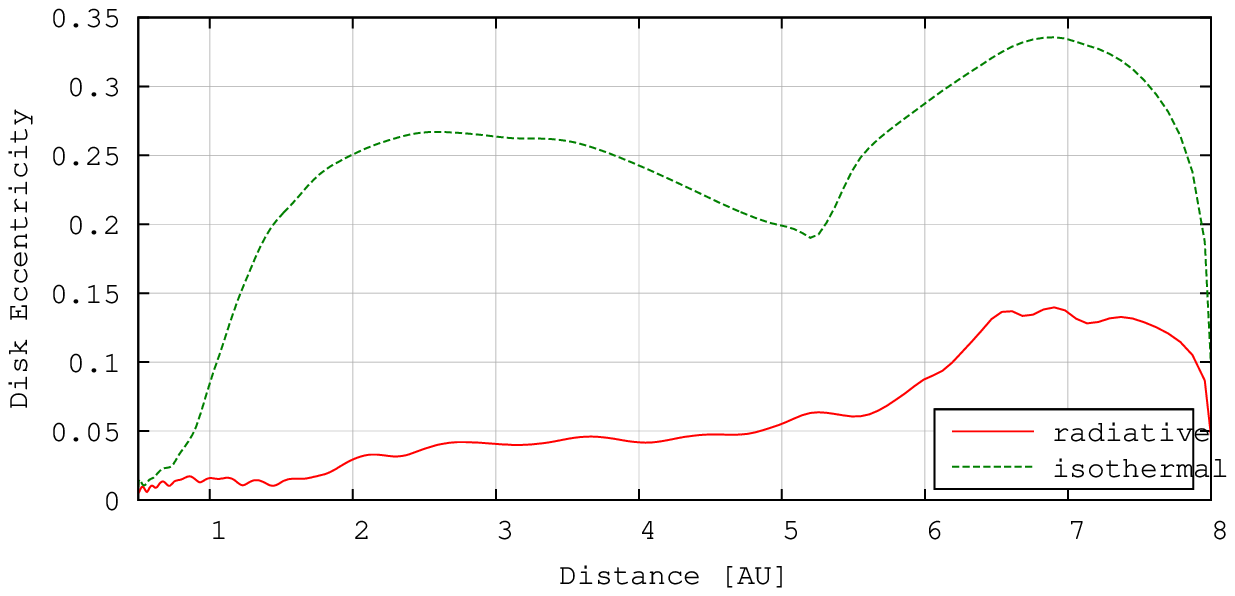} 
	\caption{
		\label{fig:gc-iso-times-sig-temp-ecc}
		Radial dependency of surface density (\textit{top}), and disk eccentricity (\textit{bottom}) of the $ H/r = 0.05 $ \textit{isothermal} model and the \textit{fully radiative} standard model after $ 750 $ binary
		orbits.
	}
\end{figure}

Now we include all physics, in particular the viscous heating and radiative cooling terms and solve the full set of equations as stated
above. For given binary parameter and disk physics (opacity and viscosity) the dynamical state and structure of the disk is 
completely determined once the disk mass has been specified. The density and temperature distribution cannot be stated freely and hence
the power law given in Table \ref{tab:standardmodel} refers only to the initial setup.

The radial disk structure is displayed in Fig.~\ref{fig:gc-times-sig-temp-ecc} where azimuthally averaged quantities of $\Sigma, T$ and $e_\mathrm{disk}$ are displayed.
The upper panel of Fig.~\ref{fig:gc-times-sig-temp-ecc} shows the surface density and temperature distribution for five different timestamps. For the initial
setup, at $ t = 0$\,years, the surface
density follows the initial $ r^{-1} $ profile in the inner region ($ 0.5$\,AU--$5$\,AU) and is lowered to the density floor in the outer region. The temperature
follows the initial $ r^{-1} $ profile for the whole computational domain. After about $25$ binary orbits a quasistationary state is reached where the profiles show 
a bend at about $ r = 1.8$\,AU. This change in slope, which occurs at a temperature of about $ 1000$\,K in the profiles, is a result of a change
in the opacity caused by the sublimation of dust grains, which starts at about
$ (\rho\,\mathrm{g}^{-1} \mathrm{cm}^3)^{1/15} \times 4.6 \times 10^3$\,K in the opacity tables (see Table~\ref{tab:opacity}) given by \citet{1985prpl.conf..981L}. A second
bend occurs at about $ r = 4.5$\,AU, which corresponds to the truncation radius of the companion's tidal forces. In the bottom panel of 
Fig.~\ref{fig:gc-times-sig-temp-ecc} the corresponding radial distribution of eccentricity is shown. The eccentricity is low in the inner
parts of the disk and increases with larger radii. The highest values for $e(r)$ occur in the outer region $ r \gtrsim 5.5$\,AU, beyond
the tidal truncation radius.
But these regions do not play an important dynamical role because there is only very little mass left owing to the tidal forces of the secondary star. 
For example, the disk has a mass-weighted mean overall disk eccentricity of $ e_{\rm disk} = 0.032 $ after $ 100 $ binary orbits even though more than half of the radial extent of the disk has an 
eccentricity of $  \geq 0.05 $. Remarkably the eccentricity is much lower than in the isothermal simulations presented in Section \ref{sec:isothermal}.
As a consequence of the lack of eccentricity there is not visible disk precession. 

Fig.~\ref{fig:gc-times-sig-temp-ecc} shows the radial dependency of the disk surface density and disk eccentricity of the $ H/r = 0.05 $ model in equilibrium after $ 750 $ binary orbits compared with the subsequently introduced
fully radiative model. The surface density in the isothermal model misses the bend caused by the $ 1000 $\,K change in the disk opacity
and is slightly steeper than the initial $ r^{-1} $ profile. 
The disk eccentricity  in the isothermal case is much more homogeneous with radius, as is to be expected for a coherent structure. This
has already been observed by \citet{2008A&A...486..617K}.

\begin{figure}
	\centering
	\includegraphics[width=\columnwidth]{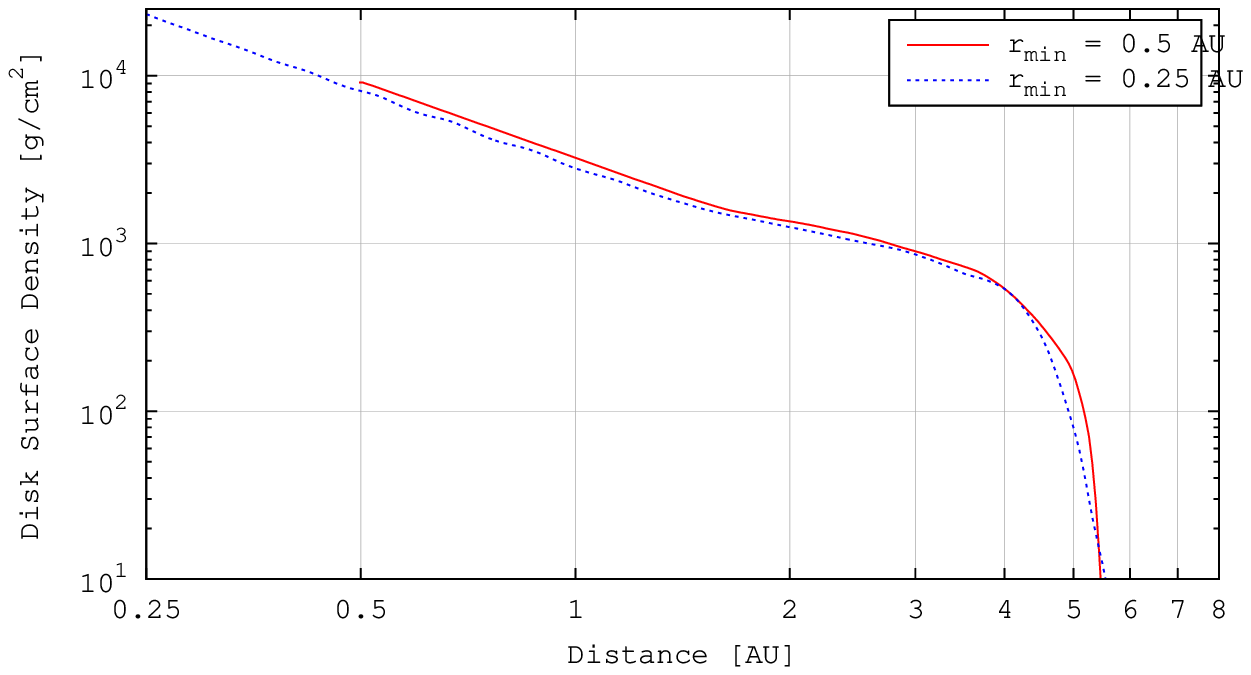} 
	\includegraphics[width=\columnwidth]{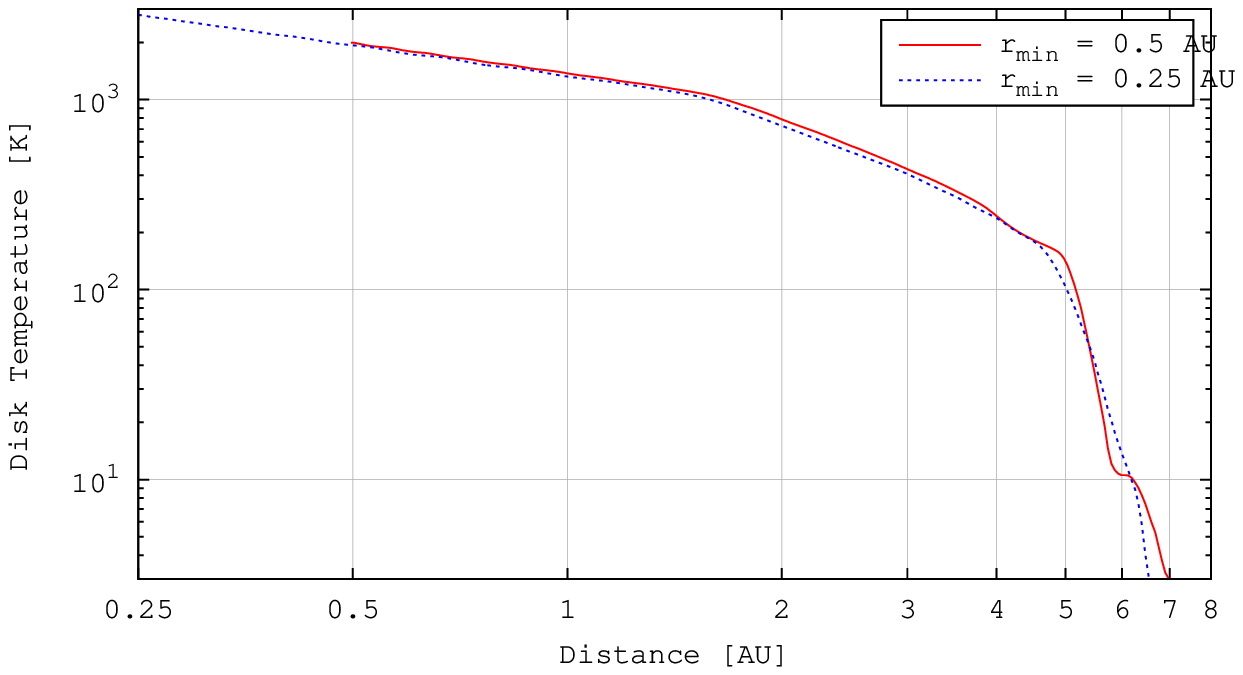}
	\caption{
		\label{fig:gc-bounds-sig-temp}
		Azimuthally averaged surface density (\textit{top}) and midplane temperature (\textit{bottom}) profiles for the radiative standard model using 
                two different locations of the inner boundary of the computational domain.
                Results are displayed at $100$ binary orbits ($6666$\,years). In the overlapping region from $0.5$--$8$\,AU the
		profiles match very well and the region from $0.25$--$0.5$\,AU in the $r_\mathrm{min}=0.25$\,AU model is a consistent continuation.
	}
\end{figure}

\begin{figure}
	\centering
	\includegraphics[width=\columnwidth]{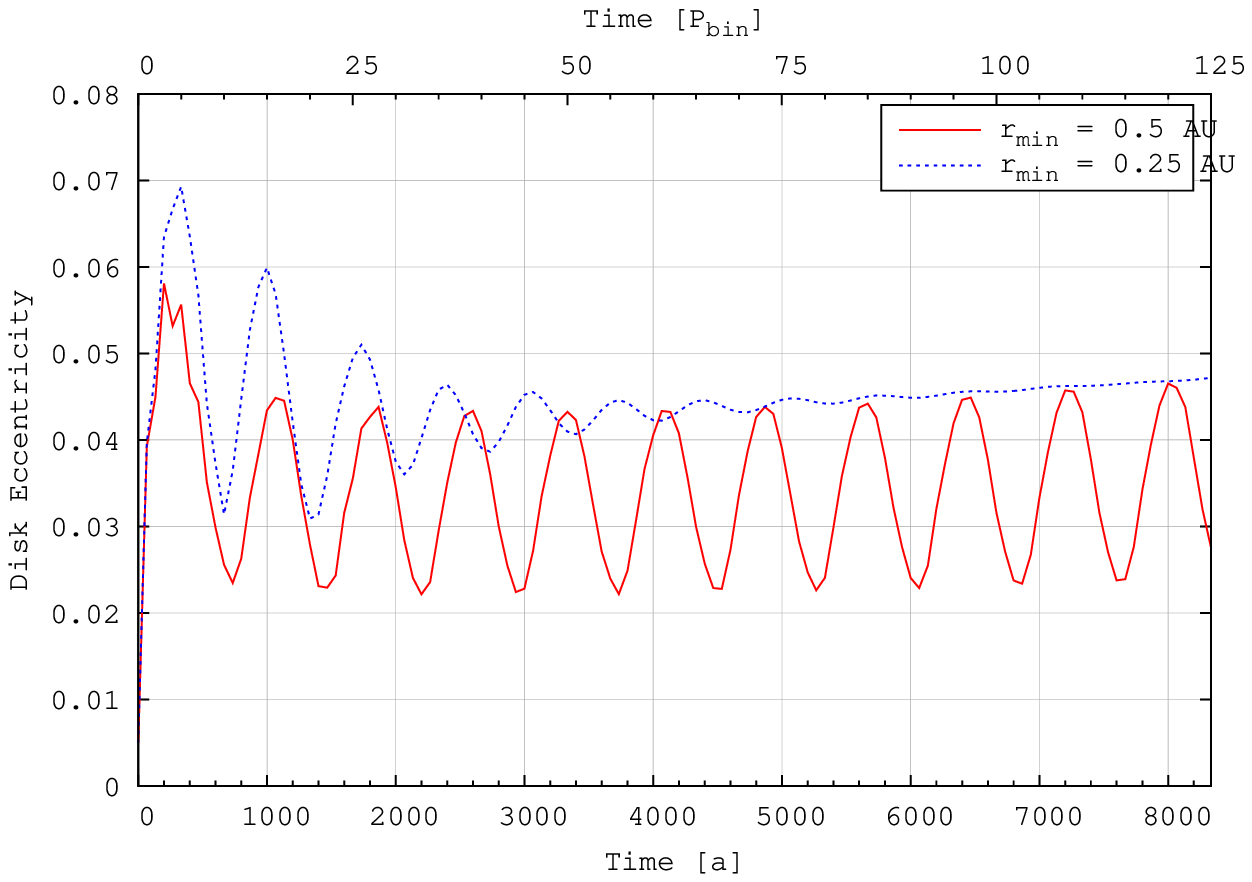} 
	\includegraphics[width=\columnwidth]{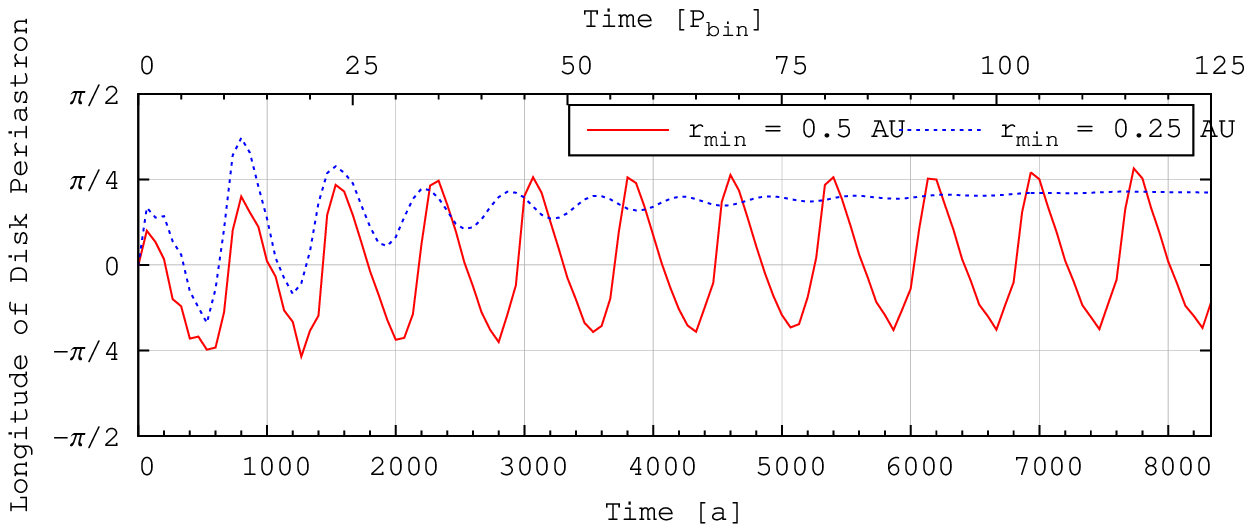}
	\caption{
		\label{fig:gc-bounds-ecc-peri}
		Global mass-weighted disk eccentricity (\textit{top}) and disk periastron (\textit{bottom}), sampled at the binary's apastron, for two different inner boundary positions.
	}
\end{figure}

To check for numerical dependencies we varied the location of the inner boundary of the computational domain and the grid resolution.
First we chose a smaller inner boundary of $r_\mathrm{min} = 0.25$\,AU increasing the number of radial grid points such that the spatial grid resolution remained
unchanged in the overlap region. In Fig.~\ref{fig:gc-bounds-sig-temp} the radial dependency of the surface density and the midplane temperature at the end of the simulations
differs not too much in the overlap region of both simulations, so that the effects of the inner boundary are negligible in this respect. However, the amplitude of the oscillation 
in the time evolution of the disk eccentricity is nearly gone (see Fig.~\ref{fig:gc-bounds-ecc-peri}). In the standard model we could see a small oscillation in the disk periastron,
which also vanishes in the $ r_\mathrm{min} = 0.25$\,AU model. The non-existence of a coherent precession is also an indication of very low disk eccentricity
in both cases.
In additional test simulations (not displayed here), we extended the outer radius to the value of $r_\mathrm{max}=12$\,AU, changing the number of grid cells accordingly to
maintain the resolution of the standard model. As expected, this larger radial domain does not change the results.

Then we varied the grid resolution by first doubling the grid cells to $512 \times 1150 $ and doubling it again to $ 1024 \times 2302 $ grid cells.
This change in resolution has basically no impact on the disk's dynamics. 
As before, the disk eccentricity settles to a time averaged value of about $ 0.04 $. In all three simulations the disk periastron 
does not reach a state with real precession and the only change caused by the change of resolution is a small shift in time. 
Consequently our standard resolution of $ 256 \times 574 $ cells seems to be sufficient to resolve the disk dynamics properly.

\section{Variation of physical parameters}

To study the influence of the physical conditions on the evolution of the disk, we investigated in particular the impact of the disk mass $ M_\mathrm{disk} $,
the viscosity $ \nu $, which is determined by $ \alpha $, the opacity $ \kappa $, and the binary's eccentricity $ e_\mathrm{bin} $. 

\subsection{Disk mass}

The first parameter we investigated is the disk mass. In contrast to isothermal simulations our radiative simulations depend on the disk mass as the opacity depends on gas density.
To analyze the influence of the
disk mass on the disk's evolution, we ran four simulations with disk masses $ M_\mathrm{disk} $ of $ 0.005\,M_{\sun}$,  $0.01 \,M_{\sun}$, $0.02\,M_{\sun}$ and $ 0.04 \,M_{\sun} $
while keeping all other parameters unchanged.

\begin{figure}
	\centering
	\includegraphics[width=\columnwidth]{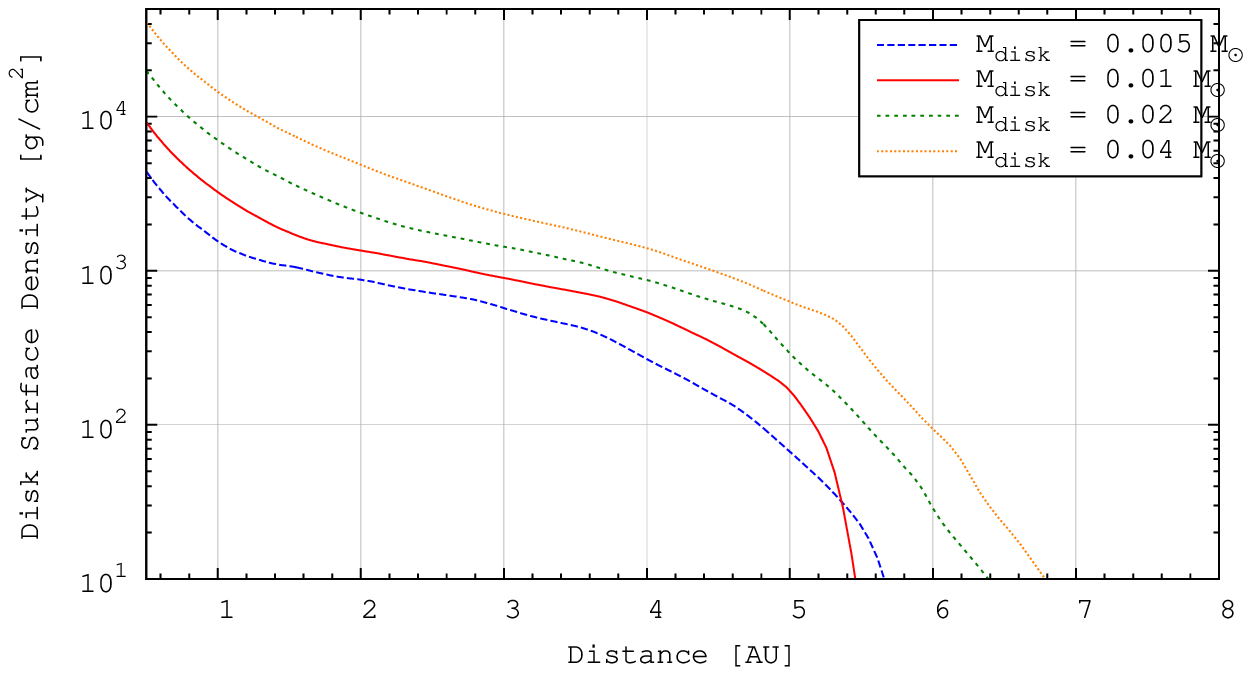} 
	\includegraphics[width=\columnwidth]{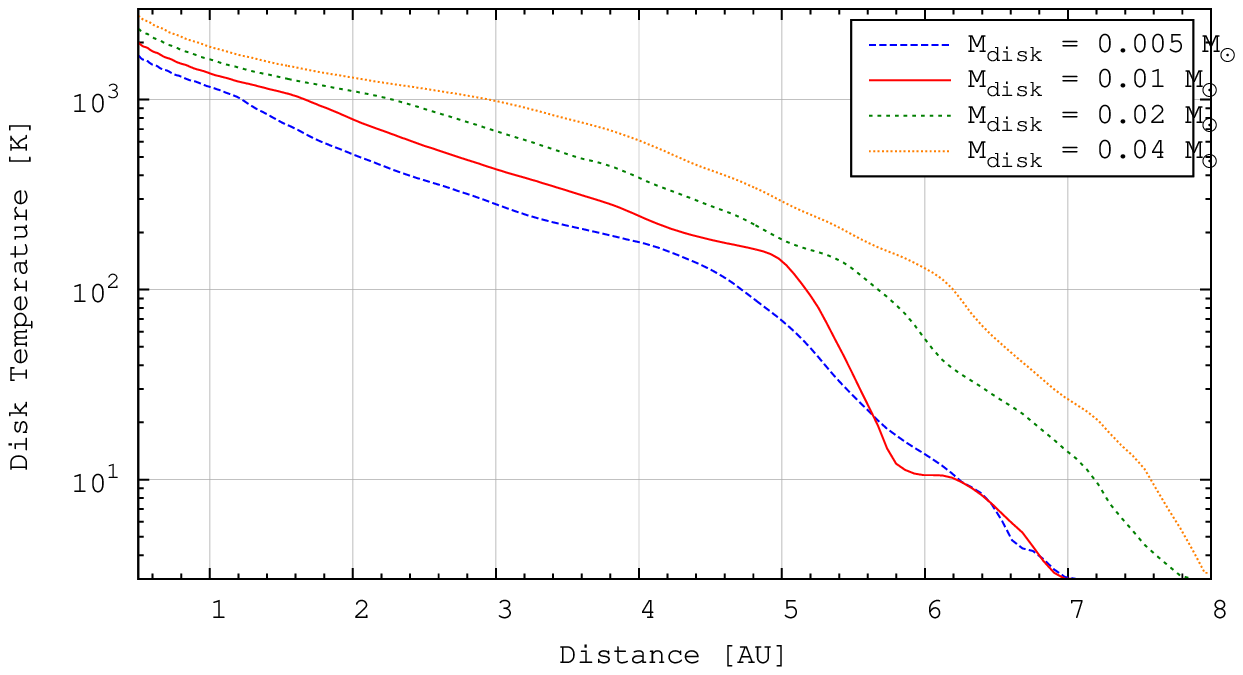}
	\caption{
		\label{fig:gc-mass-sig-temp}
		Azimuthally averaged surface density (\textit{top}) and midplane temperature (\textit{bottom}) profiles after $100$ binary orbits ($6666$\,years) for radiative
                models using different initial disk masses.
                The profiles can be fitted using power-laws when divided into two regimes: An outer region with
		temperatures of less than about $1000$~K and an inner region with temperatures above $1000$~K because of a break in the opacity tables at about $1000$~K. The
		inner region follows $ \Sigma \propto r^{-1.49 \mathrm{\ to\ } -1.51}$ and $ T \propto r^{-0.53 \mathrm{\ to\ } -0.54}$, whereas the outer region follows
		$ \Sigma \propto r^{-0.82 \mathrm{\ to\ } -1.67}$ and $ T \propto r^{-1.39 \mathrm{\ to\ } -1.58}$ for all models.
	}
\end{figure}

The surface density and temperature profiles after $100$ binary orbits for different disk masses is displayed in Fig.~\ref{fig:gc-mass-sig-temp}, where the solid (red) line refers to the standard model 
displayed in Fig.~\ref{fig:gc-times-sig-temp-ecc}. The profiles can be divided into two regimes that are separated by the $ 1000$\,K temperature line caused by the opacity (see Section \ref{sec:structureanddynamics}).
In each regime the surface density and temperature profiles follow a simple power-law that depends only weakly on the disk mass, and is given in the caption of Fig.~\ref{fig:gc-mass-sig-temp}.

\begin{figure}
	\centering
	\includegraphics[width=\columnwidth]{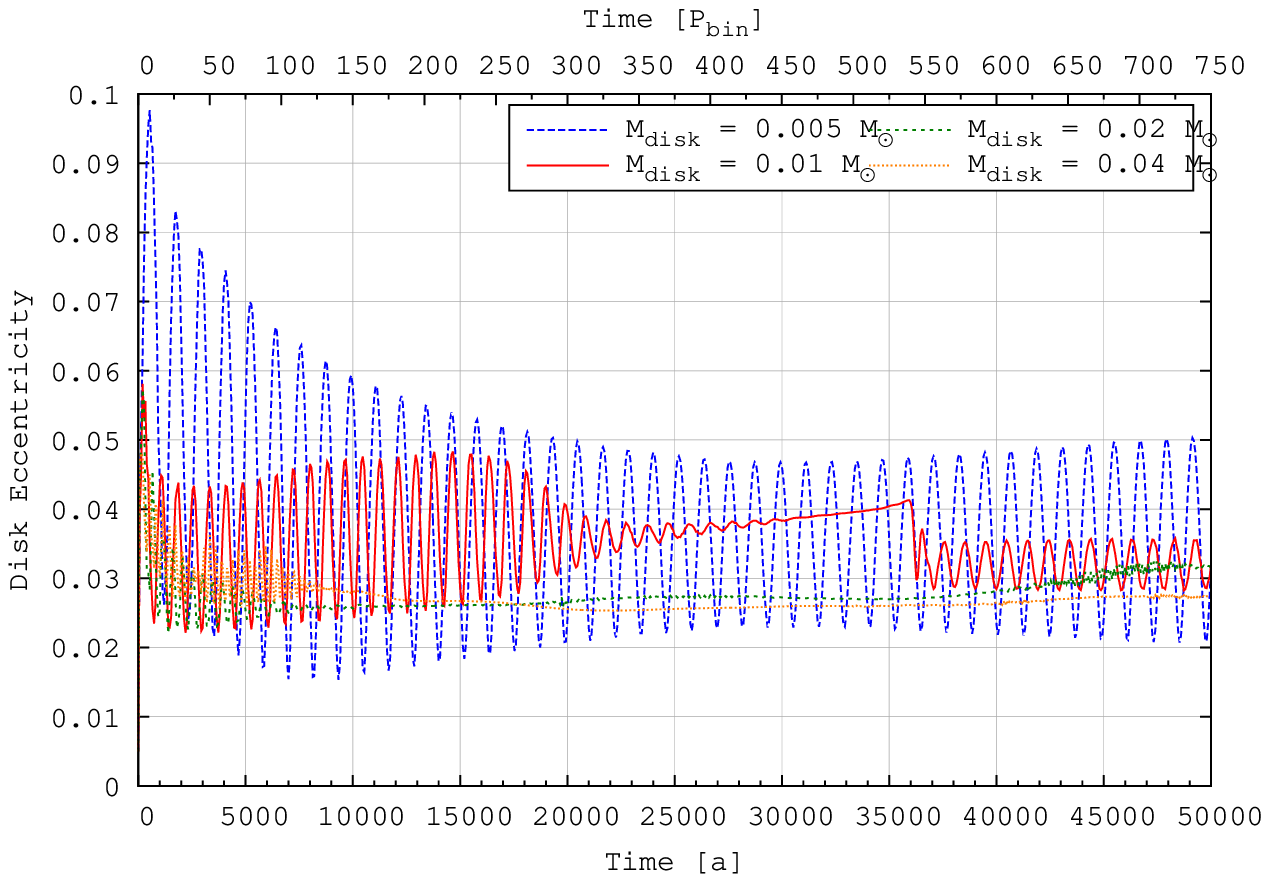} 
	\includegraphics[width=\columnwidth]{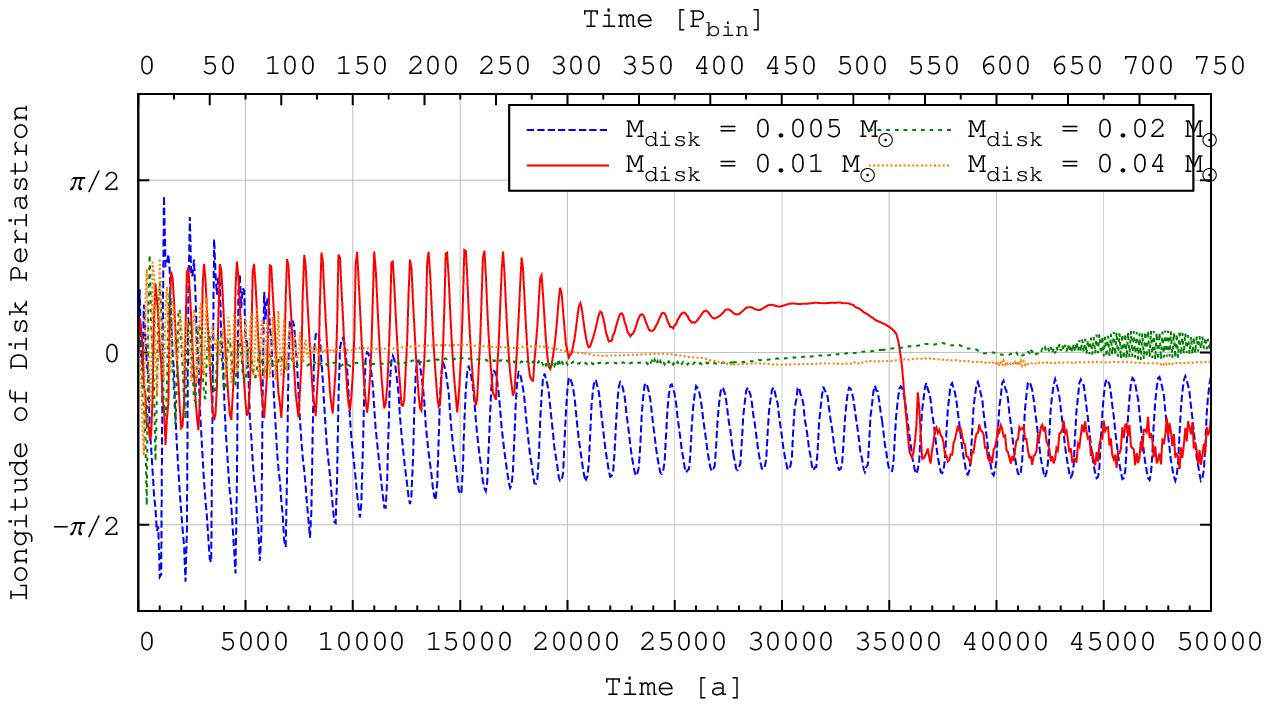}
	\caption{
		\label{fig:gc-mass-ecc-peri}
		Global mass-weighted disk eccentricity (\textit{top}) and disk periastron (\textit{bottom}), sampled at the binary's apastron, dependency for different disk masses.
	}
\end{figure}

\begin{figure}
	\centering
	\includegraphics[width=\columnwidth]{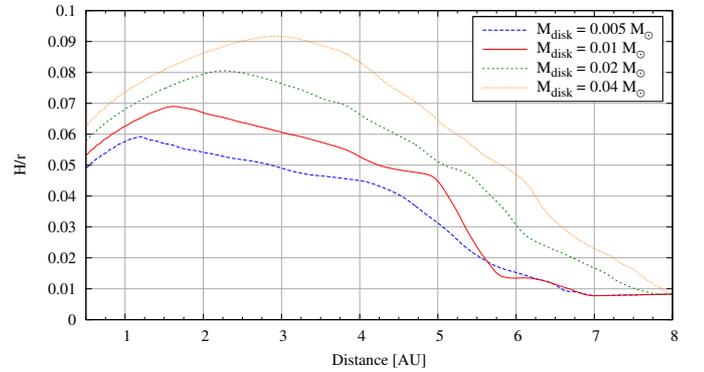} 
	\caption{
		\label{fig:gc-hr}
		$ H/r $ after $100$ binary orbits ($6666$\,years) for different disk masses.
	}
\end{figure}

Fig.~\ref{fig:gc-mass-ecc-peri} shows the time evolution of the disk eccentricity and periastron for different disk masses. 
The higher the disk mass, the lower the oscillations of the the disk eccentricity, but the time average 
of the disk eccentricity is in the range of $ 0.04 $ -- $ 0.05 $ for all disk masses. The disk periastron displays no real precession and with increasing disk mass it settles at about $ 0 $. This is
in contrast to the isothermal simulations presented in Section \ref{sec:isothermal} where we obtained a disk eccentricity $ e_\mathrm{disk} $ of $ 0.2 $ for an aspect ratio $ H/r $ of $ 0.05 $ and
a real precession of the disk periastron. There is a trend, however, that the eccentricity becomes higher for cooler disks with lower $H/r$. 
But one has to be careful here, because the aspect ratio is constant in radius and time for the isothermal simulations, but not for our radiative simulations.
Fig.~\ref{fig:gc-hr} shows the aspect ratio of the disk after $ 100 $ binary orbits. 
All models had an initial value of $ \left( H/r \right)_\mathrm{initial} = 0.05 $ but end up with different values of $ H/r $ depending on the disk mass and the phase in the binary orbit.
We note that in all models the mass of the disk reduces with time owing to the mass loss across the outer boundary. In particular,
we find that after $ 100 $ binary orbits the disks have lost about $ 27\,\% $ of their initial mass in the $ M_\mathrm{disk} = 0.04\,M_{\sun} $ model,
$ 23\,\% $ in the $ M_\mathrm{disk} = 0.04\,M_{\sun} $ model, $ 17\,\%$ in the $ M_\mathrm{disk} = 0.04\,M_{\sun} $ and $ 6\,\% $ in the $ M_\mathrm{disk} = 0.005\,M_{\sun} $ model.
Hence, the values quoted in the text and figures always refer to the initial disk masses.

The results for $e_{\rm disk}$ in Fig.~\ref{fig:gc-mass-ecc-peri} show a marginal increase of the disk eccentricity for smaller disk masses.
To test if this trend continues, we performed additional simulations for even smaller initial disk masses and found indeed
an increased oscillatory behavior of the eccentricity, which 
settles eventually after about $600$ binary orbits to a low eccentric state very similar to the $M_{\rm disk}= 0.025$ model, however.
Hence, there does not seem to exist an obvious trend of $e_{\rm disk}$ with disk mass.

\subsection{Viscosity}
\label{sec:viscosity}

In this section we investigate the influence of the viscosity $ \nu $. To study the dependence on $ \nu $ we varied the $ \alpha $ parameter, which determines the viscosity,
$ \nu = \alpha c_\mathrm{s} H = \alpha c_\mathrm{s}^2 \Omega_K^{-1} $ \citep{1973A&A....24..337S},
from our standard model ($\alpha = 0.01 $) and kept all other parameters unchanged.
We varied $ \alpha $ from $ 0.005 $ to $ 0.04 $.

\begin{figure}
	\centering
	\includegraphics[width=\columnwidth]{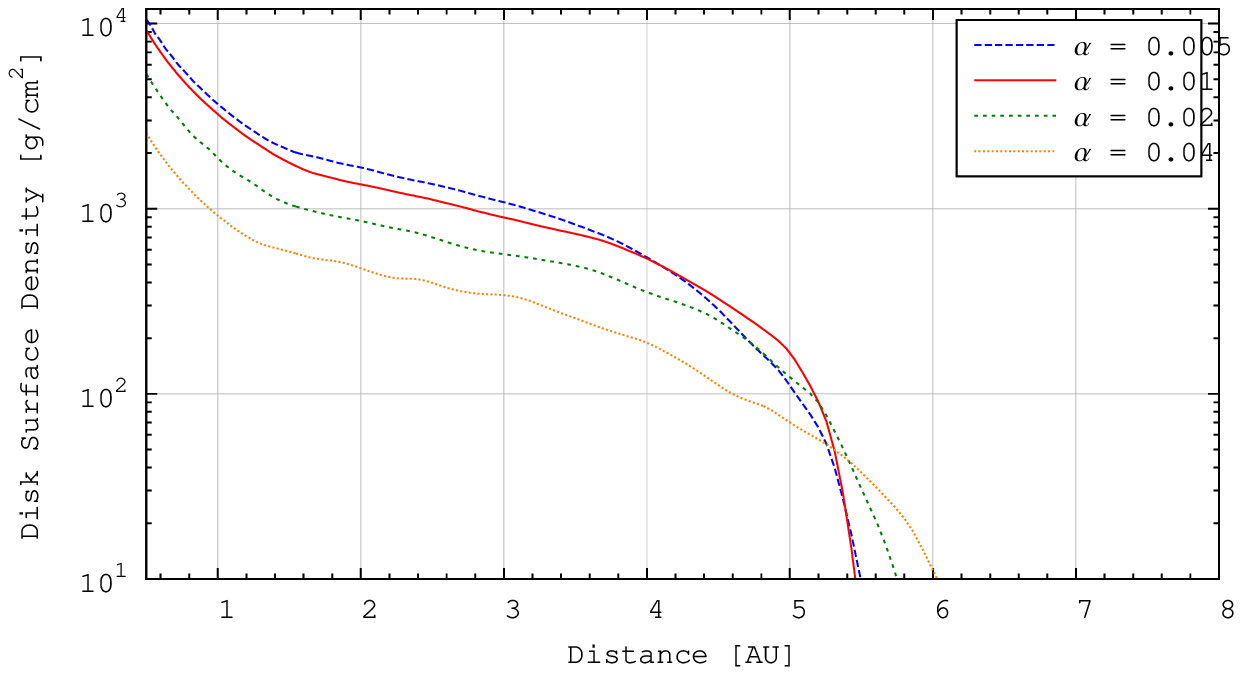} 
	\includegraphics[width=\columnwidth]{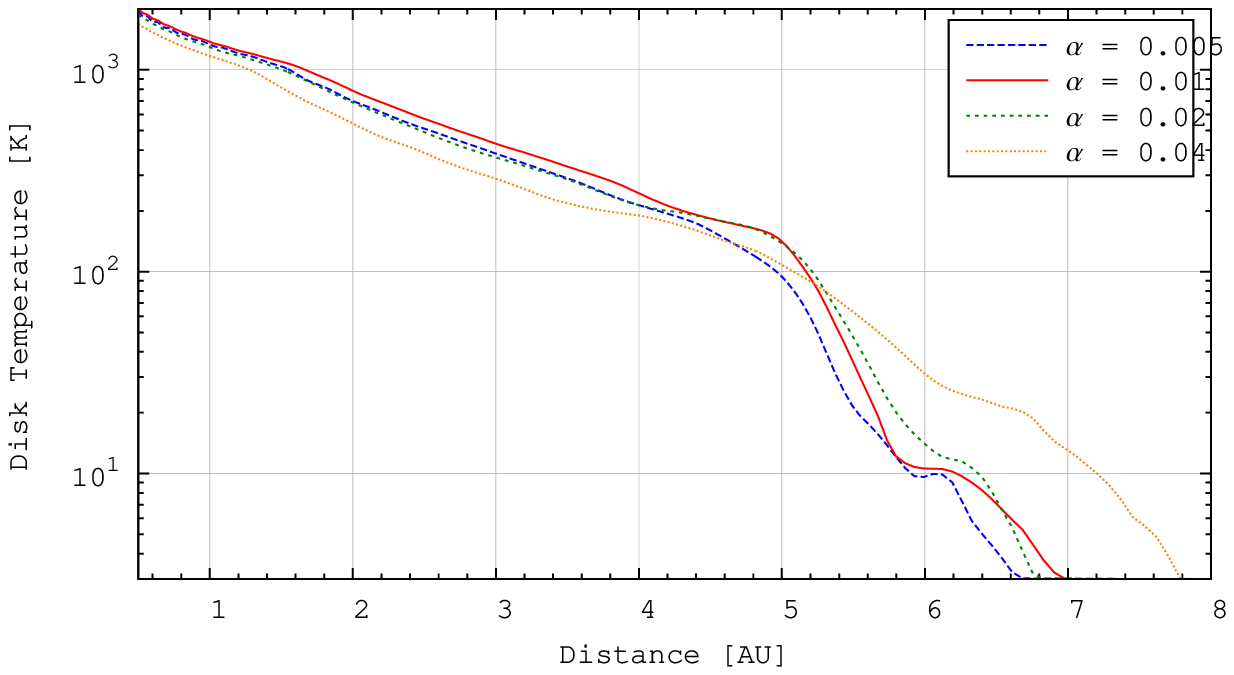}
	\caption{
		\label{fig:gc-visc-sig-temp}
		Azimuthally averaged surface density (\textit{top}) and midplane temperature (\textit{bottom}) profiles after $100$ binary orbits ($6666$\,years) 
               using four different values for the viscosity parameter $\alpha$. The profiles can be fitted using power-laws when divided into two regimes:
               An outer region with temperatures of less than
		about $1000$~K and an inner region with temperatures above $1000$~K, because there is a break in the opacity tables at about $1000$~K. The
		inner region follows $ \Sigma \propto r^{-1.48 \mathrm{\ to\ } -1.53}$ and $ T \propto r^{-0.53}$, whereas the outer region follows
		$ \Sigma \propto r^{-0.78 \mathrm{\ to\ } -0.95}$ and $ T \propto r^{-1.45 \mathrm{\ to\ } -1.53}$ for all $\alpha$ values.
	}
\end{figure}

Fig.~\ref{fig:gc-visc-sig-temp} shows the surface density and temperature profiles after $ 100 $ binary orbits.
All models started with identical disk mass, and evidently, higher $ \alpha $ models lose the
disk masses more rapidly. For example, at the displayed time at $100$ binary orbits the $ \alpha=0.04$ model has lost about
$71\,\%$ of its initial mass, while the standard model ($\alpha=0.01$) lost only $19\,\%$, see also Fig.~\ref{fig:gc-visc-mass} below.
The shape of the surface density and temperature profile seems to be independent of the viscosity in the disk, as expected. Again, the profiles can be divided into two regimes
that are separated by the $1000$\,K temperature line. The power-laws for the surface density and temperature profiles are given in the caption of Fig.~\ref{fig:gc-visc-sig-temp}.

\begin{figure}
	\centering
	\includegraphics[width=\columnwidth]{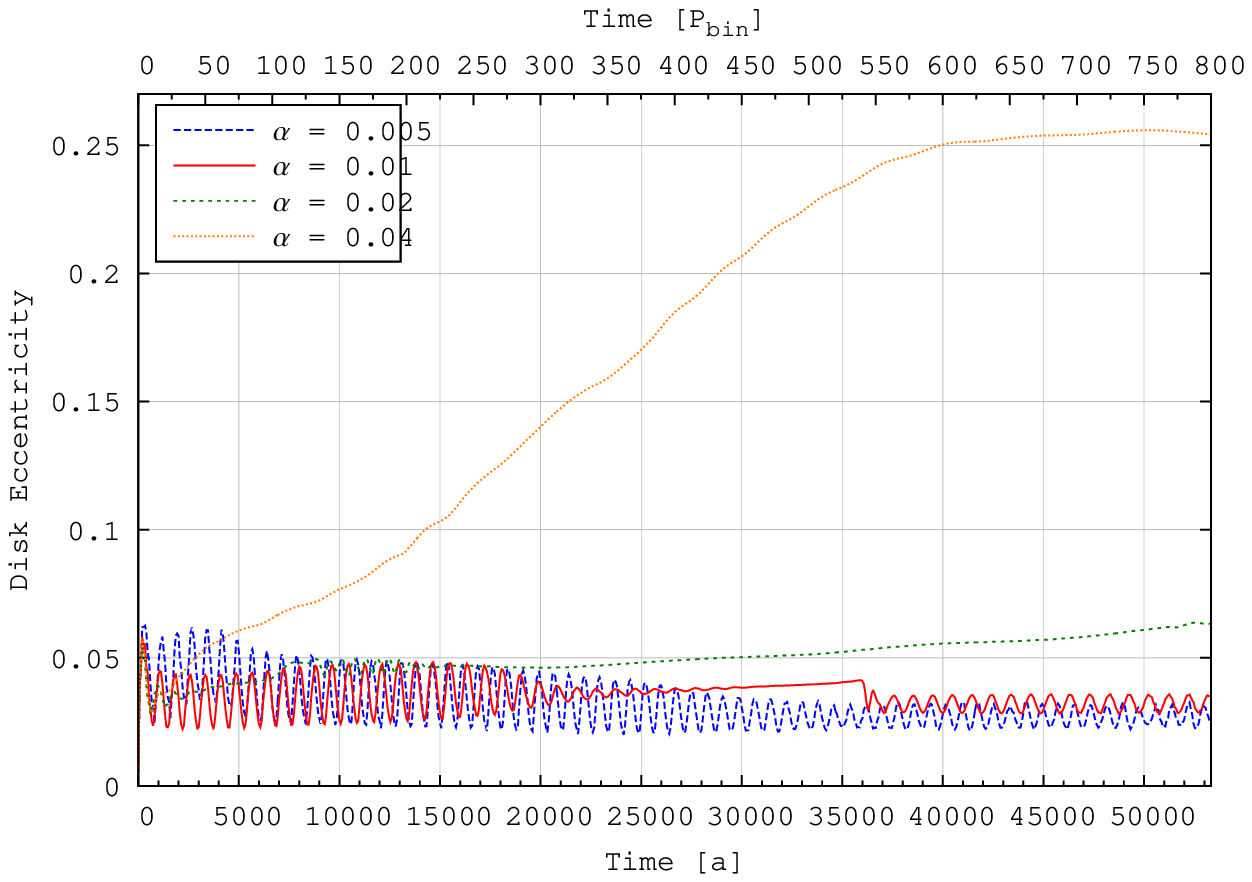} 
	\includegraphics[width=\columnwidth]{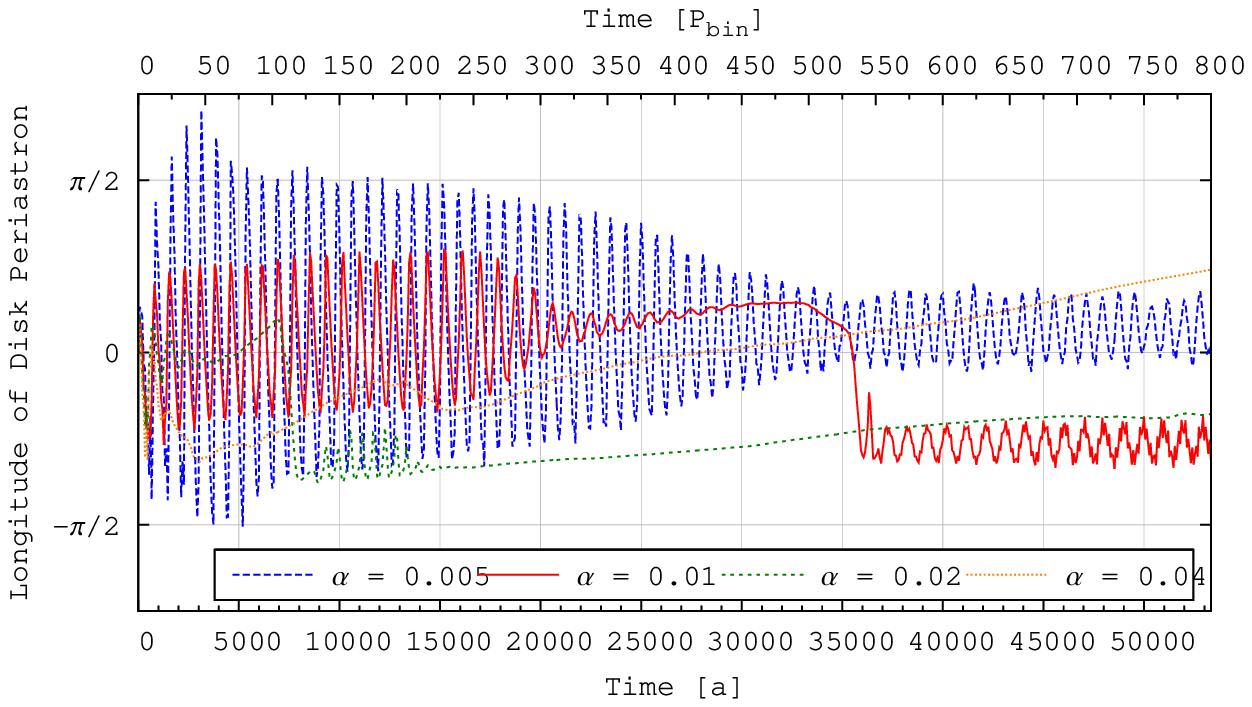}
	\caption{
		\label{fig:gc-visc-ecc-peri}
		Global mass-weighted disk eccentricity (\textit{top}) and periastron (\textit{bottom}), sampled at the binary's apastron,
                as a function of time for different $\alpha$ values for the viscosity.
	}
\end{figure}

Fig.~\ref{fig:gc-visc-ecc-peri} shows the influence of $ \alpha $ on the disk eccentricity and periastron.
Higher values of  $ \alpha $ and thus high viscosities result in a calmer disk that
does not react fast enough on the disturbances of the binary companion.
Therefore the disk shows less oscillations in the disk eccentricity and the disk periastron remains almost constant for the higher values of $ \alpha $.
The disk eccentricity in the model with the highest viscosity, $\alpha = 0.04$, increases to up to $ 0.25 $ within $ 600 $ binary orbits,
and the disk eccentricity in the $ \alpha = 0.02 $ model starts to grow slowly after about $ 200$ -- $300 $ binary orbits.
This effect of a rising disk eccentricity is a direct consequence of the increasing viscosity.
As shown in  Fig.~\ref{fig:gc-visc-mass}, the mass loss of the disk depends on the magnitude of the viscosity, the higher $\alpha$, the faster the mass
loss of the disk. For example, after $ 500 $ binary orbits the $ \alpha = 0.04 $ disk has lost about $ 97\,\% $ of its initial mass.
The increased mass loss for higher viscosities is a result of the stronger outward spreading of the disk, i.e. a larger truncation radius.
The larger disk makes the outer parts of the disk more susceptible to the tidal perturbations of the secondary, which increase the
disk eccentricity dramatically. This is in accordance with the viscosity dependence found in earlier studies, e.g. \citet{2008A&A...487..671K}.

\begin{figure}
	\centering
	\includegraphics[width=\columnwidth]{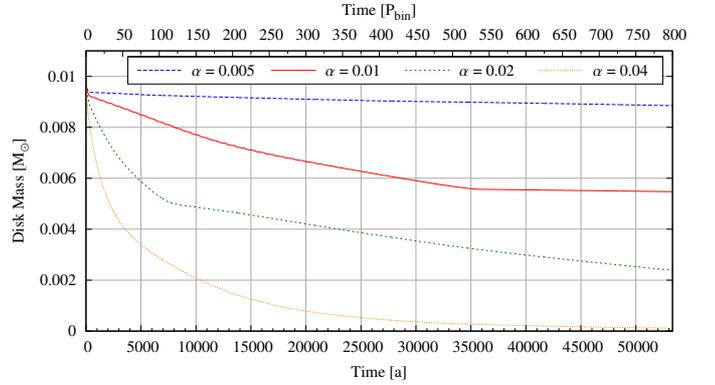}
	\caption{
		\label{fig:gc-visc-mass}
		Disk mass evolution for different $ \alpha $ values for the disk viscosity.
	}
\end{figure}

\subsection{Opacity}

The amount of cooling in our radiative model depends on the disk's opacity. To examine the influence of the opacity model we calculated the standard model with two different opacity models leaving all
other parameters unchanged. In the first calculation we used the opacity tables (see Table~\ref{tab:opacity}) shown by \citet{1985prpl.conf..981L}, whereas the second calculation uses the opacity tables given by
\citet[their Table~3]{1994ApJ...427..987B}.

As expected, the opacity plays an important role and the model with the \citeauthor{1994ApJ...427..987B} opacity
shows less oscillations in the disk eccentricity, but the time average is about the same as in the model with the \citeauthor{1985prpl.conf..981L} opacity. This is similar to the
$ r_\mathrm{min} $ change in Section~\ref{sec:structureanddynamics}. The surface density and midplane temperature profiles at the end of the simulation in the simulation
with the \citeauthor{1994ApJ...427..987B} opacity model differ slightly from the one shown in Fig.~\ref{fig:gc-times-sig-temp-ecc}. Both models have a bend at a temperature of about $1000$\,K
but the temperature profile in the model with the \citeauthor{1994ApJ...427..987B} opacity model is flatter and slightly lower in the region below $1000$\,K  and slightly steeper in the region about $ 1000$\,K.
In exchange, the surface density is slightly steeper and slightly higher in region the below $1000$\,K and matches the other model in the region above $1000$\,K very well.

\subsection{Binary eccentricity}
\label{subsec:binary-ecc}

Another very important factor for the disk's evolution are the binary parameters. 
We therefore varied the binary's eccentricity in our standard model from $ e_\mathrm{bin} = 0 $ to $ 0.4 $. Because this also changes 
the truncation radius of the disk that is caused by the binary's tidal forces \citep{1994ApJ...421..651A}, we extended the computational domain to up to $ 12.5$\,AU for the $ e_\mathrm{bin} = 0$ model. To reach the
same resolution in the computational domain of the standard model ($0.5$ -- $ 8$\,AU) we increased the number of cells in radial direction to $ 295 $ in the $ e_\mathrm{bin} = 0 $ model. The
$ e_\mathrm{bin} = 0.05 $, $ e_\mathrm{bin} = 0.1 $ and $ e_\mathrm{bin} = 0.2 $ models were adjusted accordingly in their computational domain and resolution in radial direction.

\begin{figure}
	\centering
	\includegraphics[width=\columnwidth]{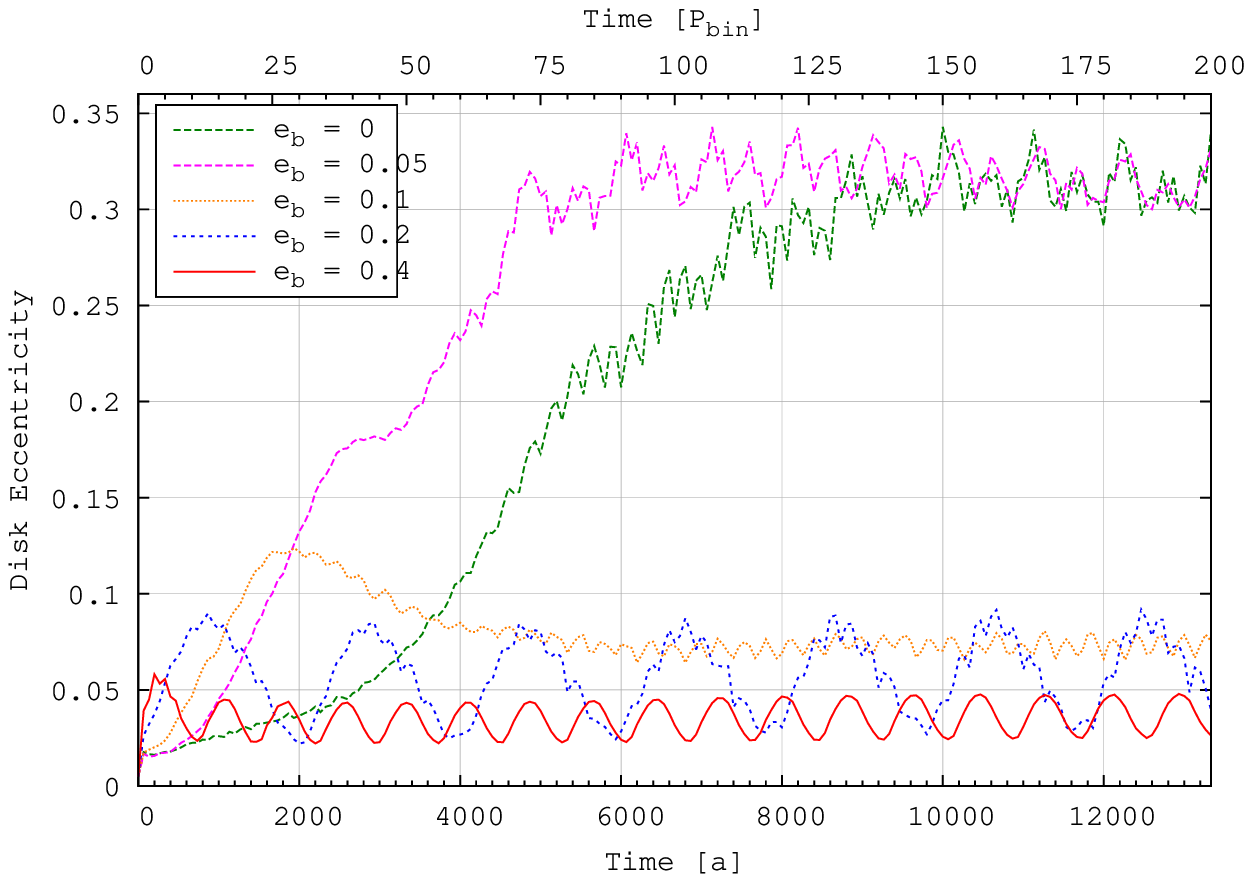} 
	\includegraphics[width=\columnwidth]{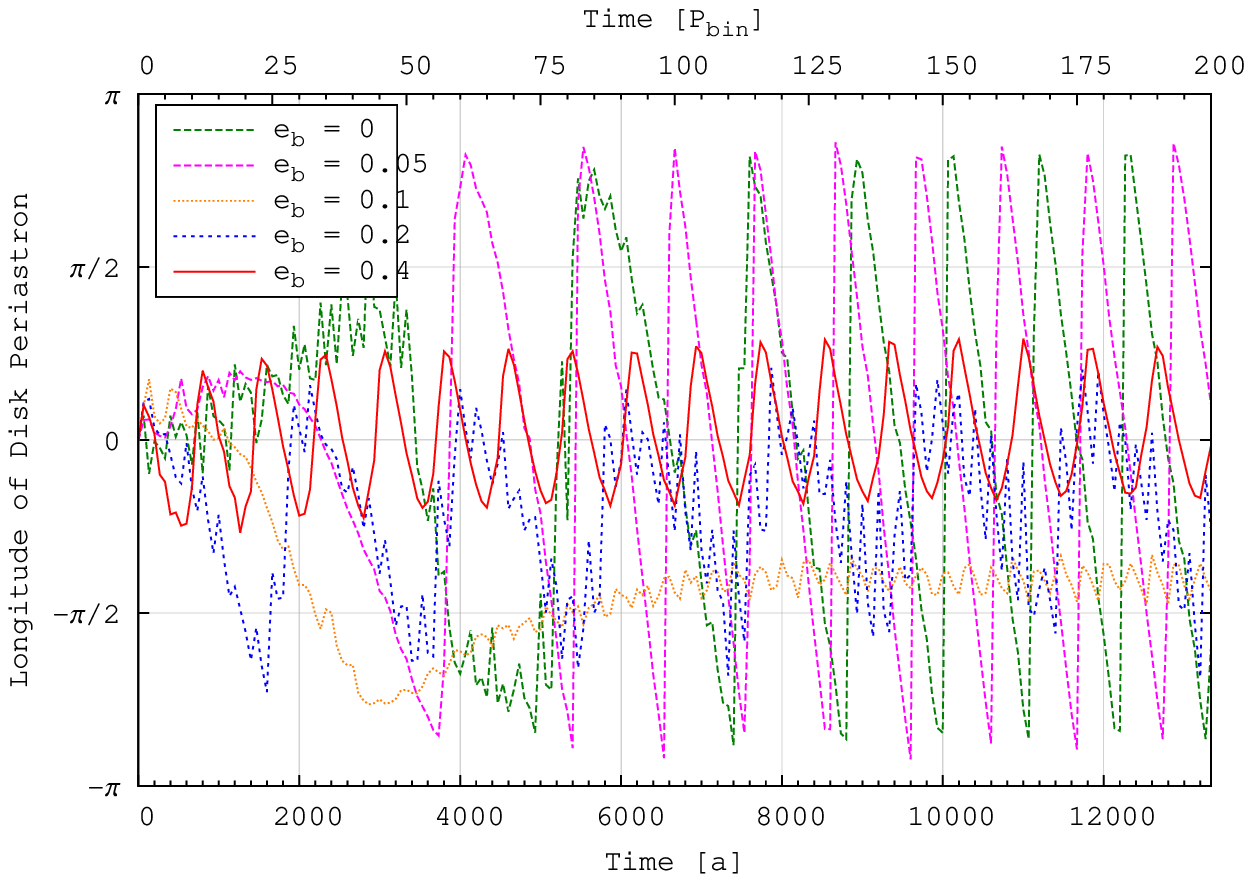}
	\caption{
		\label{fig:gc-binary-ecc-peri}
		Global mass-weighted disk eccentricity (\textit{top}) and disk periastron (\textit{bottom}), sampled at the binary's apastron, for different binary eccentricities.
                The low eccentricity ($ e_\mathrm{bin} < 0.05 $) models need much more time
		to reach their equilibrium state compared to the standard model. These are also the only models that develop a real coherent disk precession.
	}
\end{figure}

Fig.~\ref{fig:gc-binary-ecc-peri} shows the time evolution of the disk eccentricity and periastron. Interestingly, the $ e_\mathrm{bin} = 0 $ and $ e_\mathrm{bin} = 0.05 $ models show
the highest disk eccentricity. These high values for $ e_\mathrm{bin} = 0 $ seem to agree with \citet{2008A&A...487..671K}. Also, the $ e_\mathrm{bin} = 0 $
and $ e_\mathrm{bin} = 0.05 $ models are the only ones that have a real coherent disk precession with a precession rate of $ -0.033 \,P_\mathrm{bin}^{-1}$ for the $ e_\mathrm{bin} = 0 $ model and
$ -0.062\,P_\mathrm{bin}^{-1}$ for the $ e_\mathrm{bin} = 0.05 $ model. All other models with $ e_\mathrm{bin} \geq 0.1 $ show a disk eccentricity of $ e_\mathrm{disk} < 0.1 $ 
and no precession, which also indicates that the disk is not globally eccentric. For the low eccentric binaries it takes up to about $125$ binary orbits to reach the high
eccentric quasi-equilibrium disk state, which is long compared to the standard model, which reaches its quasistationary state after only about $ 15 $ binary orbits.
These timescales agree well with those obtained by \citet{2008A&A...487..671K} for isothermal disks.
The reason why binaries with low $e_{\rm bin}$ tend to have eccentric disks is the larger disk radius in this case. This
allows an easier operation of the instability according to the model of \citet{1991ApJ...381..259L}. 

The surface density and midplane temperature profiles after $100$ binary orbits of all five models have the same slope but differ slightly in absolute values. The $ e_\mathrm{bin} = 0.4 $
models is the hottest and densest model and then the temperature and surface density decreases with decreasing binary eccentricity. The $ e_\mathrm{bin} = 0 $ model shows low oscillations in the
profiles. As expected, the disk's truncation radius owing to the binary's tidal forces is shifted outward in the models with lower binary eccentricity.

\section{Brightness variations}

\begin{figure}
	\centering
	\includegraphics[width=\columnwidth]{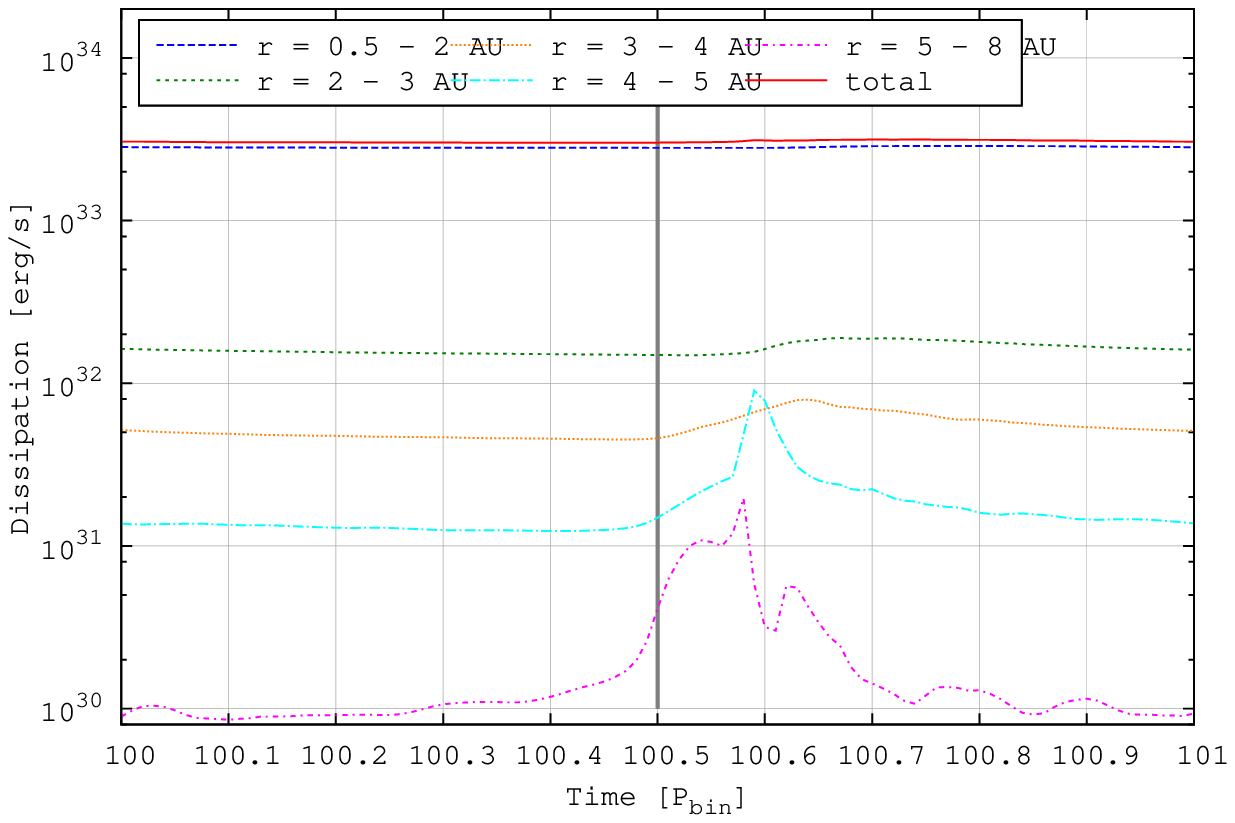} 
	\includegraphics[width=\columnwidth]{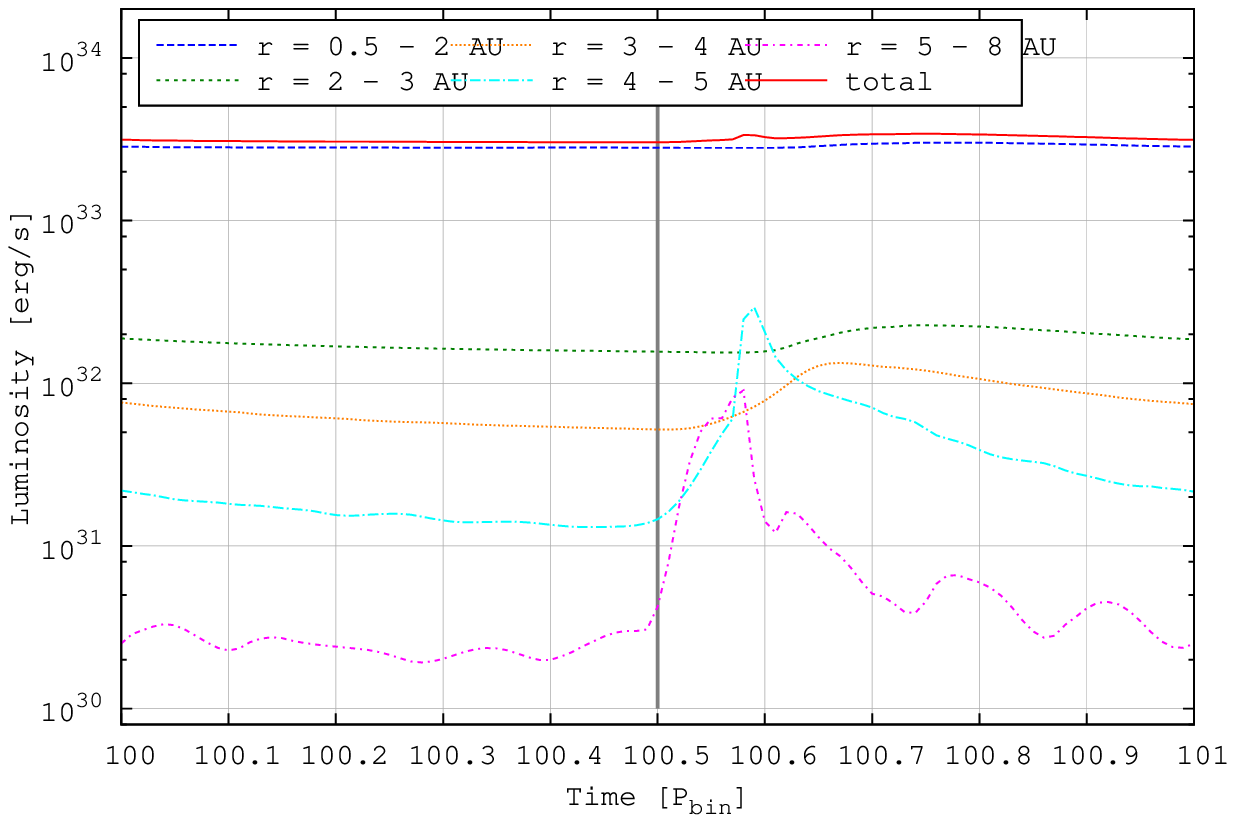} 
	\caption{
		\label{fig:gc-luminosity}
		Variation of disk dissipation and disk luminosity during one binary orbit. The dissipation and luminosity are calculated for five different rings ranging
		from $0.5$--$2$\,AU, $2$--$3$\,AU, $3$--$4$\,AU, $4$--$5$\,AU and $5$--$8$\,AU. Because the disk is truncated at about $ 4.5 $\,AU (see Fig.~\ref{fig:gc-times-sig-temp-ecc}), the
		outmost ring does not contain much mass. The solid (red) curve is the sum of all five rings. The gray line indicates the binary's periastron passage.
 	}
\end{figure}

To identify possible observable changes in the brightness of the systems caused by perturbations in the disk, we calculated 
theoretical light curves for our standard model.
For that purpose, we examined the time variation of the disk dissipation given by
\begin{eqnarray}
	D_\mathrm{disk} &= \iint Q^+ \,dA,
\end{eqnarray}
and the disk luminosity given by
\begin{eqnarray}
	L_\mathrm{disk} &= \iint Q^- \,dA = \iint 2 \sigma_\mathrm{R} \frac{T^4}{\tau_\mathrm{eff}} \,dA\,.
\end{eqnarray}
To identify the source of the brightness variations we divided the disk into five rings ranging 
from $0.5$--$2$\,AU, $2$--$3$\,AU, $3$--$4$\,AU, $4$--$5$\,AU and $5$--$8$\,AU. In
each of these rings the disk dissipation and disk luminosity was calculated. Fig.~\ref{fig:gc-luminosity} displays the variation of the disk dissipation and disk luminosity of our standard model
during one orbital orbit at $ t = 100\,P_\mathrm{bin} $ where the system has already reached its quasistationary state.
The periastron occurs at $t=100.5\,P_\mathrm{bin}$, and is indicated by the vertical line.

When the binary is at apastron, the disk is almost uniform (see Fig.~\ref{fig:gc-std-density}) and we cannot see any variations in the light curves.
However, when the binary passes periastron, it starts to perturb the outer regions of the disk and two spiral arms evolve that wind themselves
to the center of the disk. This results in a rise of the luminosity by $2$--$3$ magnitudes in the outer rings of the
disk shortly after periastron passage (see Fig.~\ref{fig:gc-luminosity}) and after a short time also in a smaller rise in the inner rings of the disk.
As the binary moves farther away from the disk, the spirals are damped out and the luminosity peak vanishes. These luminosity peaks should be observable
as a $2$-magnitude increase in the mid-infrared (MIR) because the main contributions come from the outer rings ($3$ -- $5$\,AU)
with temperatures between $ 150 $\,K and $ 450 $\,K. 

Luminosity peaks have already been observed in the mid-infrared in the T Tau S system \citep{2010A&A...517A..16V}, a binary system that is not very different from the early stage of the $\gamma$~Cephei system.
Additionally, \citeauthor{2010A&A...517A..16V} also performed some radiative simulations for the T Tau S system and pointed out that these brightness variations
could be caused by the perturbations of the binary companion. However, the brightness variations found in their disk models were very weak.

\section{$\alpha$~Centauri}

In addition to the $\gamma$~Cephei system we also performed some calculations for the $\alpha$~Centauri system, 
because it is of special interest, being the nearest star to our solar system. This system has been investigated for the possibility of planet formation, see e.g. \citet{2008MNRAS.388.1528T}.
Table \ref{tab:alphacentauri} gives an overview of the parameters of our $\alpha$~Centauri model based on \citet{2002A&A...386..280P}. We tried to keep the disk parameters the same
as in the $\gamma$~Cephei model, so that the main difference are the mass ratio $ q = M_\mathrm{secondary} / M_\mathrm{primary} $ and the binary's orbital parameters.
In the $\gamma$~Cephei model we have a mass ratio $ q $ of  $ 0.28 $, an eccentricity $ e_\mathrm{bin} $ of $ 0.4 $ and a semi-major axis $ a $ of $ 20$\,AU, whereas
in the $\alpha$~Centauri model we have a mass ratio $ q $ of $ 0.84 $, an eccentricity $ e_\mathrm{bin} $ of $ 0.52 $ and a semi-major axis $ a $ of $ 23.4$\,AU.

\begin{figure}
	\centering
	\includegraphics[width=\columnwidth]{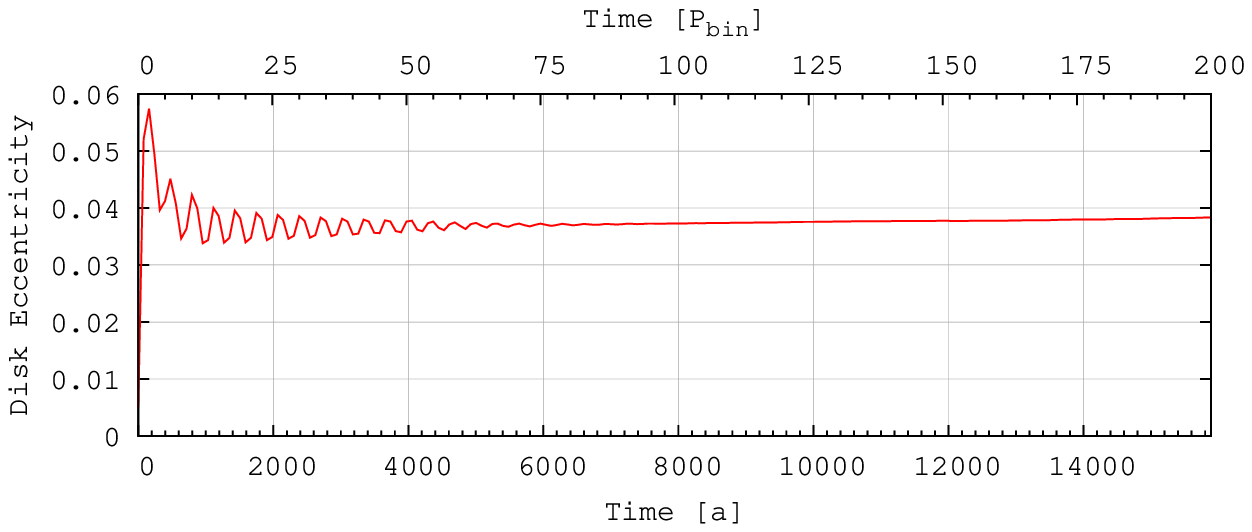} 
	\includegraphics[width=\columnwidth]{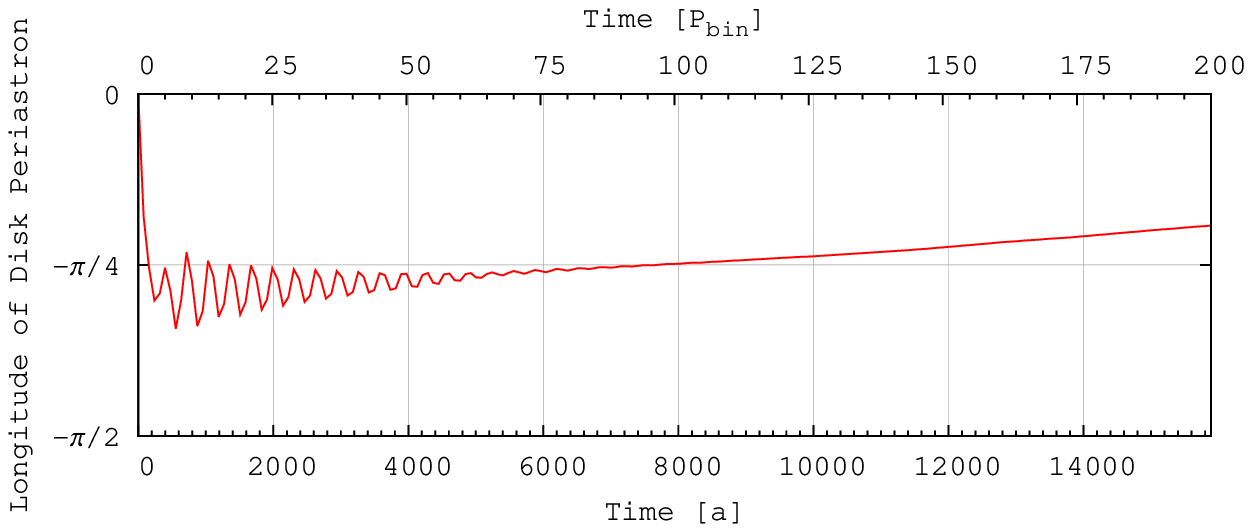}
	\caption{
		\label{fig:alp-cen-ecc-peri}
		Global mass-weighted disk eccentricity (\textit{top}) and disk periastron (\textit{bottom}), sampled at the binary's apastron for the $\alpha$~Centauri model.
	}
\end{figure}

In Section \ref{subsec:binary-ecc} we saw that for high binary eccentricities the disk eccentricity does not reach very high values, so that we would expect a fairly low
disk eccentricity for the $\alpha$~Centauri system. \citet{2008A&A...487..671K} showed that for isothermal simulations the disk eccentricity in the quasistationary state
does not depend heavily on the mass ratio $ q $. Fig.~\ref{fig:alp-cen-ecc-peri} shows the evolution of the disk eccentricity and periastron over time. The disk eccentricity
settles after about $15$ binary orbits at a rather low value of about $ 0.038 $ and the disk periastron is also at a nearly constant position.

The surface density and midplane temperature profiles for the $\alpha$~Centauri model can only be fitted in the inner region with $ r < 1.8$\,AU as a simple power-law. The surface density can be described by
a $ r^{-1.49} $ and the midplane temperature by a $ r^{-0.53} $ power-law. In the outer region the surface density and temperature decreases rather fast to $ 0 $ until about $4$\,AU, where the disk is
truncated by the binary's tidal forces.

\begin{table}
	\caption{
		\label{tab:alphacentauri}
		Parameters of the $\alpha$~Centauri model. The top entries refer to the fixed binary parameter. Below we list along with the
		disk properties the initial disk setup and the computational parameters.
	}
	\centering
	\renewcommand\arraystretch{1.2}
	\begin{tabular}{ll}
		\hline 
		Primary star mass ($ M_\mathrm{primary} $) & $ 1.1\,M_{\sun} $ \\
		Secondary star mass ($ M_\mathrm{secondary} $) & $ 0.93\,M_{\sun} $ \\
		Binary semi-major axis ($ a $) & $ 23.4\,\mathrm{AU} $ \\
		Binary eccentricity ($ e_\mathrm{bin} $) & $ 0.52 $ \\ 
		Binary orbital period ($ P_\mathrm{bin} $) & $ 79.4431\,\mathrm{a} $ \\
              \hline
		Disk mass ($ M_\mathrm{disk} $) & $ 0.01\,M_{\sun} $ \\
		Viscosity ($ \alpha $) & $ 0.01 $ \\
		Adiabatic index ($ \gamma $) & $ 7/5 $ \\
		Mean-molecular weight ($ \mu $) & $ 2.35 $ \\
              \hline
		Initial density profile ($ \Sigma $) & $ \propto r^{-1} $ \\
		Initial temperature profile ($ T $) & $ \propto r^{-1} $ \\
		Initial disk aspect ratio ($ H/r $) & $ 0.05 $ \\
              \hline
		Grid ($N_r \times N_\varphi$) & $ 256 \times 574 $ \\
		Computational domain ($ R_\mathrm{min} $ -- $ R_\mathrm{max} $) & $0.5$ -- $8\,\mathrm{AU} $ \\
		\hline
	\end{tabular}
\end{table}

\section{Summary and conclusions}
We have investigated the dynamics of a protostellar disk in binary star systems using specifically the orbital
parameter of $\gamma$~Cephei and $\alpha$~Centauri. 
We assumed a coplanar system and used a two-dimensional hydrodynamical code to evolve the non-self-gravitating disk.
Extending previous simulations, we included internal viscous heating
given by an $\alpha$-type viscosity prescription, and radiative cooling from the disk surface.

In a first set of simulations we investigated locally \textit{isothermal} disks for different disk temperatures. 
We showed that disks in binaries of the $\gamma$~Cephei type with a standard thickness of $H/r = 0.05$ become eccentric 
($e_{\rm disk} \approx 0.2$) showing a coherent disk precession. This agrees well with previous simulations by \citet{2008A&A...486..617K} and \citet{2008MNRAS.386..973P}.
Varying the temperature in the disk, we showed that the magnitude of the disk's eccentricity becomes lower when the disk
thickness increases. For disks with $H/r \gtrsim 0.065$ the mean average eccentricity has dropped below 0.08 and 
the disks do not show a precession anymore.

Then we studied more realistic disks with internal heating and radiative cooling, varying the disk's mass.
In all cases we found relatively low eccentricities and no precession. 
We attribute the lack of eccentricities firstly to the increased disk height, which is,
in particular for the more massive disks, higher than the standard value (see Fig.~\ref{fig:gc-hr}).
Secondly, in the full radiative models the disk's dynamical behavior is more adiabatic compared to the
locally isothermal case. Then, through compressional heating ($p dV$-work), kinetic energy is transferred to internal energy,
which leads to a reduced growth of disk eccentricity. We have checked that purely adiabatic models show an even lower disk
eccentricity than the radiative models. Hence, the radiative case lies between the adiabatic and isothermal, as expected.

Because the disk's energy balance is determined via the viscosity, we changed the value of the parameter
$\alpha$ ranging from $0.005$ to $0.04$, all values that are consistent with the results of MHD-turbulent accretion
disks. Here, we found that only the disk with the highest $\alpha$ becomes eccentric. The reason for this rise is the
larger disk radius, which leads to an enhancement of the tidal torques from the secondary.
We note that the disk's outer radius in our models still lies well inside the 3:1 resonance with the binary. 
According to the linear instability model by \citet{1991ApJ...381..259L}, the disk eccentricity is excited through the 3:1 resonance
and hence, the disk should be sufficiently large, a condition which is fulfilled only for small mass ratios, $q=M_\mathrm{secondary}/M_\mathrm{primary}$.
However, as shown by \citet{2008A&A...486..617K}, disks in binary star systems with large mass ratios can turn eccentric as well, even
though the disks are small, a feature confirmed in our simulations.

The inferred short lifetime of disks with standard viscosities is slightly alarming with respect to
planet formation in these systems. In the core-accretion scenario planet formation proceeds along a sequence of many steps
that take a few Myr. For disks to persist this long in $\gamma$~Cephei-type binaries a very low viscosity
of $\alpha \lesssim 10^{-4}$ seems to be required. In the gravitational instability scenario the timescale for
planet formation is much shorter and hence, this scenario may be favored by our findings.
Observationally, several recent studies indeed suggest that the lifetime of disks in young binary stars
is significantly reduced compared to disks around single stars 
\citep{2009ApJ...696L..84C,2010ApJ...709L.114D,2011arXiv1109.4141K}. Dynamically, this behavior is expected, because the
perturbation of the companion star leads quite naturally to an increased mass loss of the disk, details
depending on the binary separation and eccentricity. 

A very critical phase in the core accretion scenario, in particular in binary stars,
is the initial growth of meter to km-sized planetesimals. Here, the growth depends on the successful sticking of
the two collision partners. Since the relative velocity of the bodies is increased in binary stars,
planetary growth will be significantly hindered by the presence of a companion,
see e.g. \citet{2006Icar..183..193T,2011CeMDA.tmp...40T} and references therein.
Here our results indicate that planetesimal growth is less negatively influenced because the disk eccentricity
is reduced for more realistic radiative disks. As has been shown, eccentric disks tend to increase the
mutual relative velocities of embedded objects, in particular of different sizes, because of the misaligned
periastrons of the particles \citep{2007arXiv0705.3421K,2008MNRAS.386..973P}.  
Hence, a radiative disk with a low viscosity could help to promote planetesimal growth.
However, it remains to be seen how the inclusion of stellar irradiation (from both stars) influences the dynamics.
Owing to the additional heating of the disk, we expect even more mass loss from the system and possibly higher
disk eccentricities because the disks are more isothermal and will have a larger radius.

Previous studies have indicated that a in mutually inclined system planetesimal growth may be enhanced because planetesimals
can be size-sorted in differently inclined planes \citep{2009ApJ...698.2066X,2009A&A...507..505M}.
However, recently \citet{2011A&A...528A..40F} showed through full 3D hydrodynamical studies that for inclined
binaries the relative velocities of different sized planetesimals increases through inclinations effects. They conclude
that for inclined systems planetesimal formations can take place only for very distant binary stars with
$a_{\rm bin} \gtrsim 60$~AU.
However, the simulations considered only isothermal disks, and it remains to be seen how radiative effects influence
the disk. But full 3D radiative simulations are still beyond the present computational possibilities, because
thousands of orbital timescales of the disk will have to be calculated.

\begin{acknowledgements}
	We would like to thank Markus Gyergyovits for useful discussions.
	Tobias M\"uller received financial support from the Carl-Zeiss-Stiftung. Most of the simulations were performed
	on the bwGRiD cluster in T\"ubingen, which is funded by the Ministry for Education and Research of Germany and
	the Ministry for Science, Research and Arts of the state Baden-W\"urttemberg. Additionally, we used
        the cluster of the Forschergruppe FOR 759 "The Formation of Planets: The Critical First Growth Phase" funded by
	the German Research Society (DFG).
\end{acknowledgements}

\bibliographystyle{aa}
\bibliography{mueller2011}

\end{document}